\documentclass[aoas,MSNbibl,nameyear,dvips]{arximspdf}
\usepackage{graphicx}

\doi{10.1214/11-AOAS475}
\volume{5}
\issue{3}
\pubyear{2011}
\firstpage{1839}
\lastpage{1875}

\makeatletter

\def\bptnote#1{}

\renewcommand{\epsilon}{\varepsilon}
\makeatother

\begin{document}
\begin{frontmatter}

\title{Modeling item--item similarities for personalized recommendations on Yahoo! front page}
\runtitle{Modeling item--item similarities}

\begin{aug}
\author[A]{\fnms{Deepak} \snm{Agarwal}\ead[label=e1]{dagarwal@yahoo-inc.com}},
\author[A]{\fnms{Liang} \snm{Zhang}\corref{}\ead[label=e2]{liangzha@yahoo-inc.com}}
\and
\author[B]{\fnms{Rahul} \snm{Mazumder} \ead[label=e3]{rahulm@stanford.edu}}
\runauthor{D. Agarwal, L. Zhang and R. Mazumder}
\affiliation{Yahoo! Labs, Yahoo! Labs and Stanford University}
\address[A]{D. Agarwal\\
L. Zhang\\
4401 Great America Pkwy\\
Santa Clara, California 95054\\
USA\\
\printead{e1}\\
\hphantom{\textsc{E-mail}:\ }\printead*{e2}} %adresu isvedimo
%komanda gale!
\address[B]{R. Mazumder\\
Department of Statistics\\
Stanford University\\
Stanford, California 94305\\
USA\\
\printead{e3}}
\end{aug}

% HISTORY:
\received{\smonth{5} \syear{2010}}
\revised{\smonth{4} \syear{2011}}

% ABSTRACT
%
\begin{abstract}
We consider the problem of algorithmically recommending items to users
on a~Yahoo! front page module. Our approach is based on a~novel
multilevel hierarchical model that we refer to as a User Profile Model
with Graphical Lasso (UPG). The UPG provides a personalized
recommendation to users by simultaneously incorporating both user
covariates and historical user interactions with items in a model
based way. In fact, we build a per-item regression model based on a
rich set of user covariates and estimate individual user affinity to
items by introducing a latent random vector for each user. The vector
random effects are assumed to be drawn from a prior with a precision
matrix that measures residual partial associations among items. To
ensure better estimates of a precision matrix in high-dimensions, the
matrix elements are constrained through a Lasso penalty. Our model is
fitted through a penalized-quasi likelihood procedure coupled with a
scalable EM algorithm. We employ several computational strategies like
multi-threading, conjugate gradients and heavily exploit problem structure
to scale our computations in the E-step. For the M-step we take recourse
to a scalable variant of the Graphical Lasso algorithm for covariance selection.

Through extensive experiments on a new data set obtained
from Yahoo! front page and a benchmark data set from a movie
recommender application, we show that our UPG model significantly
improves performance compared to several state-of-the-art methods in
the literature, especially those based on a bilinear random effects
model (BIRE). In particular, we show that the gains of UPG are
significant compared to BIRE when the number of users is large and the
number of items to select from is small. For large item sets and relatively
small user sets the results of UPG and BIRE are comparable. The UPG leads to
faster model building and produces outputs which are interpretable.
\end{abstract}

% KEYWORDS
%
\begin{keyword}
\kwd{Recommender systems}
\kwd{collaborative filtering}
\kwd{matrix factorization}
\kwd{item--item similarities}
\kwd{graphical lasso}.
\end{keyword}

\end{frontmatter}

%s1 ###
\section{Introduction}
Selecting items for display on a web page to engage users is a
fundamental problem in content recommendation
[\citet{agarwalnips}]. Item selection is made to maximize some utility
of interest to the publisher. For instance, a news site may display
articles to maximize the total number of clicks over a long time
horizon. For each display, feedback obtained from user-item
interaction is used to improve item selection for subsequent
visits. At an abstract level, the problem of recommending items on
some module of a web page can be described as follows:
\begin{itemize}
\item A user visits a web page. Typically, covariates like demographic
information, geographic location, browse behavior and feedback from
previous user visits are available for users.
\item A serving scheme selects item(s) to display on a small number of
slots in the module. The number of available slots are generally
smaller than the number of items to choose from. Typically, item(s)
selection is based on scores computed through a statistical model.
\item The user interacts with items displayed on the module and
provides feedback (e.g., click or no-click).
\item Based on feedback, parameter estimates of statistical models are
updated. The latency of update
(e.g., 5 minutes, 30 minutes, 1 day) depends on the statistical model,
the delay in receiving feedback from a user visit and the
engineering infrastructure available.
\item The process of serving items is repeated for every user visit. On
portals like Yahoo!,
there are hundreds of millions of daily visits.
\end{itemize}

%s1.1 ###
\subsection{Background and literature}\label{seccontext}
The item recommendation problem described above is closely related to
a rich literature on recommender systems and collaborative filtering
[\citet{recommsurvey}], a proper survey of which is beyond the
scope of
this paper. We describe some popular approaches that are closely
related to methods proposed in this paper.

Recommender systems are algorithms that model user-item interactions to provide
personalized item recommendations that will suit the user's taste.
Broadly speaking,
two types of methods are used in such systems---\textit{content
based} and
\textit{collaborative filtering}.\footnote{The
term ``collaborative filtering'' was coined by developers of the first
recommender system,
Tapestry [\citet{tapestry}].}
Content based approaches model interactions through user and item covariates.
Collaborative filtering (CF), on the other hand, refers to a set of
techniques that model user-item interactions
based on user's past response alone, no covariates are used. Modern day
recommender
systems on the web tend to use a hybrid approach that combines content
based and collaborative filtering.

A popular class of methods in CF are based on item--item and/or user-user
similarities [\citet{sarwar2001}; \citet{wang2006}]. These\vadjust{\eject}
are nearest-neighbor
methods where the response for a user-item
pair is predicted based on a local neighborhood average.
In general, neighborhoods are based on similarities between
items/users that are estimated through correlation measures like
Pearson, cosine similarity and
others. A better approach to estimate similarities has also been
recently proposed in \citet{koren2010factor}.\looseness=-1

Nearest neighbor methods have been used extensively in large-scale commercial
systems [\citet{linden03}; \citet{nag08}].
However, item--item similarities are measured in
terms of \textit{marginal} correlations and do not adjust for the
effect of
other items. It is not trivial to incorporate both covariates and past
responses in a
principled way. Also, the algorithms do not have a probabilistic
interpretation, which makes it
difficult to get estimates of uncertainty.
We address these issues in this paper by working in a model based framework.
A crucial aspect of our approach is in explicitly
incorporating item--item interactions after adjusting for covariates.
In fact, we model partial associations among items; it provides more
flexibility compared to
the classical item--item similarity approach that only exploits marginal
associations.
This leads to significant improvement in performance as illustrated
in Section~\ref{secexp}.

Research in CF received a boost
after Netflix ran a challenge on a movie recommendation problem.
The task was to use 100M ratings provided by half a million users on
roughly 18K movies
to minimize out-of-sample RMSE on a test set [\citet{bell2007chasing}].
The publicly available data set released by Netflix
does not contain any user or item covariates, hence,
prediction using CF is a natural approach. Several methods were
tried; the winning entry was an ensemble of
about $800$ models. Significant improvements in accuracy were attributable
to a few methods. A new class of methods that were based on SVD style
matrix factorization provided
excellent performance and were significantly better than classical
neighborhood based approaches in CF.
These are bilinear random effects models
that capture user-item interactions through a multiplicative
random-effects model.
[See
\citet{koren07}; \citet{bennett2007netflix}; \citeauthor{salakhutdinov2008bayesian}
(\citeyear{salakhutdinov2008bayesian,salakhutdinov2008probabilistic})
for more details.]
Recently, these bilinear random effects models were
generalized to simultaneously account for both covariates and past
ratings [\citet{rlfm}]. We shall refer to this class of models as
\textit{BIRE} (bilinear-random effects model) in the rest of the paper.
Methods proposed in this paper are compared to \textit{BIRE} in the
experimental section
(Section~\ref{secexp}) along with a theoretical analysis of how
the approach proposed in this paper is related to \textit{BIRE}
(Section~\ref{secbireupg}).
Through empirical analysis, we find that our approach has significantly
better predictive accuracy than \textit{BIRE}
when the number of users is large and item set to recommend from is
small; for a large sized item set and a relatively small
user set the performance is comparable to \textit{BIRE}. Indeed,
for a~large item set we find the predictions from both \textit{BIRE} and
our model to be similar.\looseness=-1

In terms of deployed large scale recommender systems, there is
published work
describing some aspects of the Amazon system
based on item--item similarity [\citet{linden03}]. In \citet
{das2007google},
techniques that power recommendations on Google News are
described. They are primarily based on item--item similarity and
Probabilistic Latent Semantic Indexing (PLSI). We compare both these methods
with ours in Section~\ref{secexp}. A~large body of work in computational
advertising [\citet{broder2008computational}] that recommends ads
to users
is also an example of recommender
problems. Most existing papers in this area focus on estimating
click-rates on
ads by users in a given context. Early work focused mostly on
covariate based regression [\citet
{chakrabarty08}; \citet{ewa}]. Recently,
\citet{agarwal2010estimating} describe an approach that combines
covariate based
regression with publisher-ad interaction through a~multilevel
hierarchical model. Item--item similarities in this model are
estimated by exploiting a known hierarchical clustering on the item space
that is obtained from domain knowledge. No such knowledge is
available in our scenario, hence, the methods described in that paper
do not apply. A new and emerging scientific discipline called content
optimization [\citet{agarwalnips}] that aims at recommending appropriate
content for a user visit to a web page is another example of a
popular recommender problem. In fact, the motivating application on Yahoo!
front page we describe in this paper is an instance of content optimization.

We also note that recommender systems in general are complex and involve
simultaneous optimization of several aspects, some of these are not
necessarily statistical. For instance, in constructing an item
pool to recommend from, human editors on a web portal like Yahoo! may
discard a Lady Gaga story if it is not compatible with the Yahoo!
brand, even if it is likely to click well. In computational advertising
on search engines, ads that
are not topically relevant to a query are removed from the item set to
begin with.
Nevertheless, statistical models that
estimate the propensity to respond positively when an item is displayed
to a user in a given context are
integral to the success of most modern day recommender systems. Data
obtained from such systems
consist of many categorical variables like user, item, URL, IP address,
search query and many more. It is
typical for such categorical
variables to have a large number of levels; new levels appear routinely
over time and the distribution of
data is heavy-tailed (a few levels are popular and a large number have
small sample size). Furthermore,
the modeling involves estimating interactions among several such
categorical attributes; data sparseness
due to high dimensionaity and imbalance in sample size is a major issue
when fitting such statistical models.
The modeling approach described in this paper provides a possible solution.

%f1 ###
\begin{figure}

\includegraphics{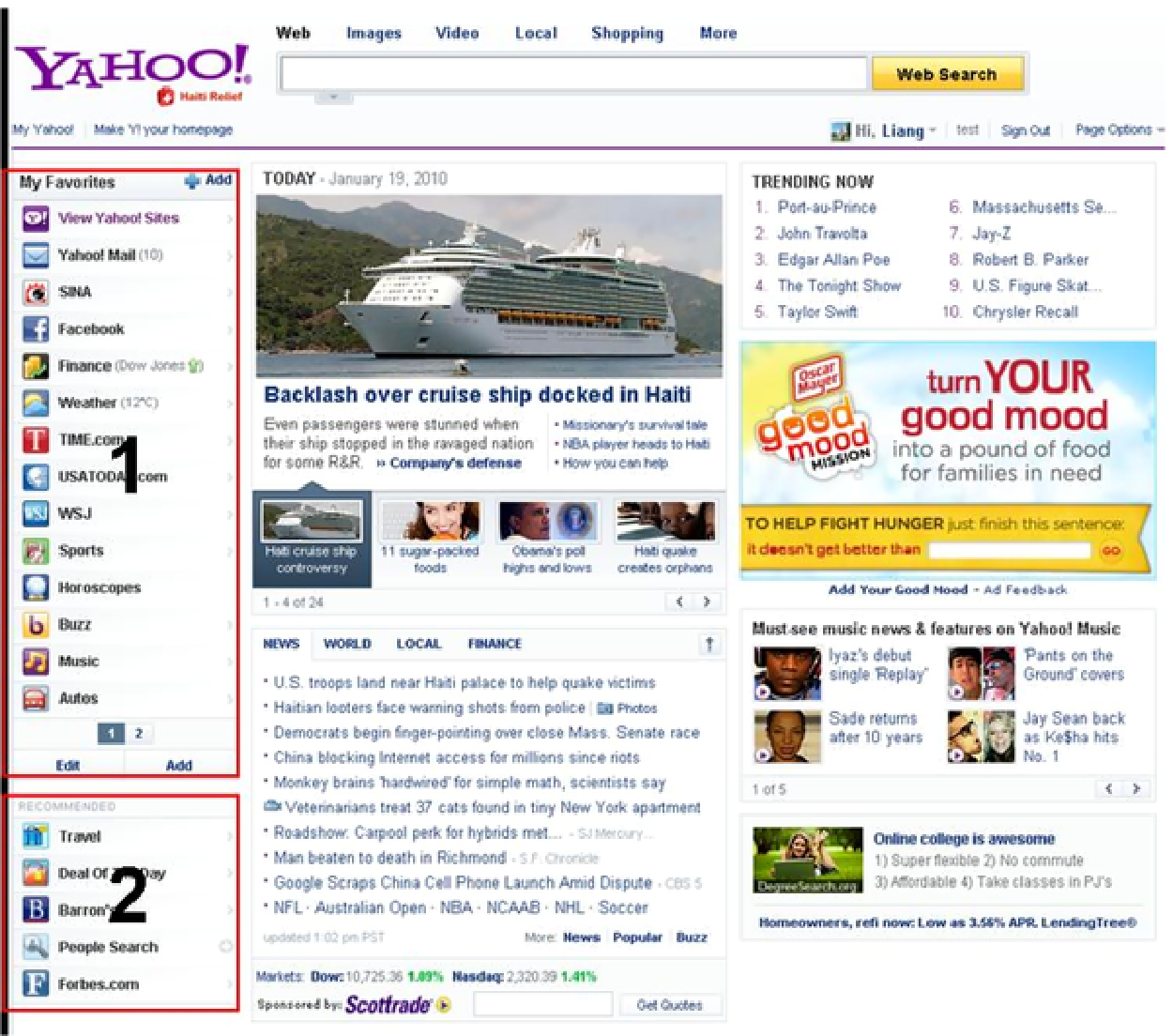}

\caption{The Personal Assistant (PA) Module on Yahoo! front page.
Region 1 displays items selected through editorial oversight and user
add and remove, Region 2
has a couple of slots where items recommended through statistical
methods are placed.}
\label{figfp}
\end{figure}

%s1.2 ###
\subsection{Motivating application}\label{secmotivate}
\begin{figure}

\includegraphics{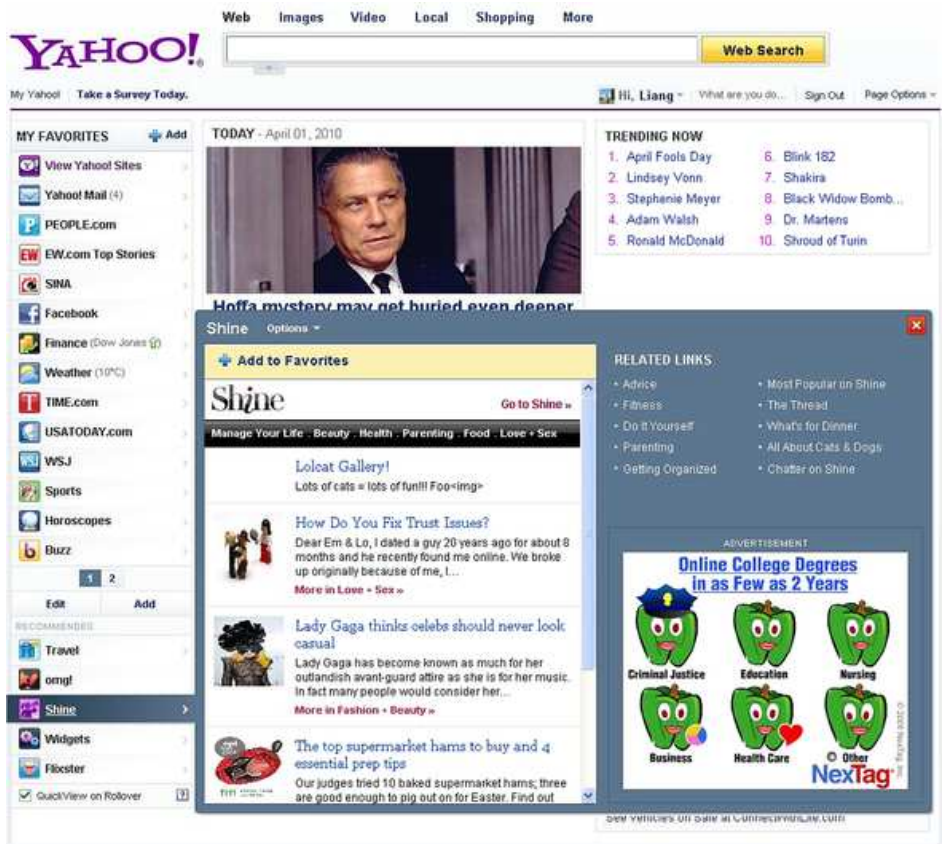}

\caption{In the ``quickview'' mode, the pop-up window when a user
hovered on one PA item ``Shine'' (\protect\url{http://shine.yahoo.com/}).}
\label{figfp2}
\end{figure}

Our motivating application requires recommending items on a Yahoo!
front page (\url{http://www.yahoo.com}) module.\vadjust{\eject}
Figure~\ref{figfp} shows the location of our front page module that
we shall refer to as the Personal Assistant (PA) module.
The items to recommend
on the PA module could consist of web apps, RSS feeds and even
websites. Some examples of the PA items include
Gmail,\footnote{\url{http://mail.google.com/}.}
Facebook,\footnote{\url{http://www.facebook.com/}.} Yahoo!
Travel,\footnote{\url{http://travel.yahoo.com/}.} Yahoo!
Games,\footnote{\url{http://games.yahoo.com/}.}
CNN\footnote{\url{http://www.cnn.com/}.} and so on. For
instance, the Gmail web app enables login to Gmail
from the Yahoo! front page. The PA module is composed
of two regions---the upper part of the module (region 1) consists of
items that have been added by the user and the lower part (region 2)
consists of items that are recommended. There are
four ways a user can interact with PA items---hover, click, add and
remove. ``Hover'' only works in ``quickview'' mode; Figure~\ref{figfp2}
shows an interaction with a PA item when it is hovered upon. For some
PA items,
a click redirects to the corresponding
website. If a user likes some recommended item displayed in region 2, it
can be added to region 1 by clicking on the ``Add'' button. A user can
also remove items from region 1. To simplify the problem, we treat
both ``hover'' and ``click'' as positive feedbacks of similar strength in
our models (henceforth, both
are referred to as ``click'').

%f2 ###

Our main focus is to
recommend items for slots in region 2 to maximize the overall
click-rate on the PA module. At the time of writing this paper, items in
region 1 were preselected through editorial oversight. A user may,
however, decide to remove some of these and add new ones. Out of five
slots in region~2, three showed editorially selected items while the
other two (the second and the third positions) display items
recommended through statistical methods. Although we anticipate items
being algorithmically recommended on the entire PA module in the
future, for the sake of illustration we only focus on recommending
items for the
second and the third positions in region 2.

%s1.3 ###
\subsection{Statistical challenges}
Successful deployment of large scale recommender systems like PA involves
several statistical challenges. Providing
personalized recommendations is important since item affinity is often user
specific. However, the frequency distribution of user visits to Yahoo! front
page is skewed; a~small fraction of users are frequent visitors, while
the remaining are \textit{tourists} with infrequent visits. Hence, the
sample size available to estimate item affinity per user is small for
a large number of users. Informative covariates
are often available for users based on demographic information
(through registration data), geo-location (based on IP-address) and
inferred browse behavior (based on historical user activities throughout
the Yahoo! network). For users with small sample sizes, estimating item
affinity through covariates is an attractive strategy. The statistical
challenge is to build a model that provides user-specific item
affinity for heavy users but falls back on covariate based estimates in
small sample size scenarios. We propose a novel multilevel
hierarchical random effects model to perform such estimation.

Multilevel hierarchical random effects models are well studied in the
statistics literature [see \citet{gelman2007data} for a review].
Simple versions
provide an attractive framework to perform small sample size
corrections when the number of replications have large variation
across groups.
%The corrections work by the well understood ``borrowing
%of strength'' phenomenon. In some applications like spatial statistics
%and time series, a single random effect is associated with each
%observation. Smoothing in this case occurs by assuming a joint
%distribution on all random effects with dependencies modeled through
%spatial and/or temporal proximity.
However, in our scenario the number of random effects far exceeds the
number of data points. We have a vector of user random effects that
represents latent user affinity to the entire item set, but a typical
user interacts with a small number of items. This gives rise to a
large missing data problem---the full latent affinity vector for
each user needs to be estimated by using partial user response
data. Model fitting in such scenarios presents additional challenges
that we address in this paper. In particular, we constrain the random
effects through a prior that models latent item dependencies using a
well studied notion in the recommender systems literature---item--item
similarities. But unlike existing methods in recommender
problems that estimate similarities through marginal correlations, we
incorporate such item--item similarities in a model based way through
partial correlations.

We also discuss how to perform fast online
updates of model parameters. Online parameter update makes the model
adaptive and provides
better recommendations in practice [\citet{agarwalfobmkdd10}].
Also, a~website like Yahoo! front page receives
hundreds of millions of visits on a~daily basis, hence, scalability
of the model fitting procedure is an important consideration.
We scale our computations by taking recourse to the penalized quasi-likelihood
procedure (PQL) for model fitting [\citet{breslowclayton}]. Our prior
involves estimating
a high-dimensional covariance matrix; we discuss methods to perform
such estimation in a
scalable way for large problems with thousands of items.

%s1.4 ###
\subsection{Overview of our proposed modeling approach}
The item pool to recommend from in PA is small (approximately 50), and
it is also hard to obtain informative covariates for the items
themselves. Hence, we build a per-item regression model (\textit{IReg})
to estimate the odds of a click on an item for a given user visit.
Thus, if $\mathbf{x}_{u}^{(t)}$ denotes the covariate
vector for user $u$ at time $t$, the log-odds $\theta_{uj}^{(t)}$ of a
click when item $j$
is displayed to user $u$ at time $t$ is modeled as
$\theta_{uj}^{(t)}=\mathbf{x}_{u}^{(t)'}\bolds{\beta}_{j}$. The\vspace*{+2pt} coefficient
vector $\bolds{\beta}_{j}$ for each item $j$ is estimated through a
logistic regression with a ridge penalty on the coefficients.

Although the per-item user covariate logistic regression model IReg
is satisfactory for users with a moderate to small number of
visits, it may not be the best for frequent visitors where it is
attractive to have a model that exploits response from previous user
visits. Thus, it is desirable to have a model that smoothly
transitions from IReg to a per user model depending on the sample
size. We accomplish this by augmenting our regression model with
user-specific\vspace*{-1pt} random effects. In other words, we assume
$\theta_{uj}^{(t)} = \mathbf{x}_{u}^{(t)'}\bolds{\beta}_{j} +
\phi_{uj}^{(t)}$, where user $u$ has a\vspace*{-2pt} random vector
$\bolds{\phi}_{u}^{(t)} = (\phi_{u1}^{(t)},\ldots,\phi_{uJ}^{(t)})$ ($J$
is the number of items). The estimated log-odds is now based on both
regression and user-specific random-effects. As in all random-effects
models, one has to perform smoothing by imposing an appropriate
prior. We assume $\bolds{\phi}_{u}^{(t)}$ i.i.d. $\sim
\operatorname{MVN}(\mathbf{0},\bolds{\Sigma})$, where the prior precision matrix
$\bolds{\Omega} = \bolds{\Sigma}^{-1}$ is estimated from the data. The prior
precision matrix measures partial associations between item pairs after
adjusting for the covariate effects. This ensures we are more likely
to recommend items that are positively correlated to items that the
user liked in the past. For instance, if users who like
PEOPLE.com\footnote{\url{http://www.people.com}.} also like
EW.com,\footnote{\url{http://www.ew.com/ew}.} a new user who visits
PEOPLE.com will be recommended EW.com.

Most of the problems that we are interested in are
under-determined---number of observations being
relatively small compared to the complexity of our covariance model.
This naturally calls for a
regularization, for good predictive accuracy.
Though there are several possibilities for the regularization of the
(inverse) item covariance,
for the context of this paper we resort to a~structure encoding
conditional independencies among items. We achieve this via
a sparse inverse covariance regularization for the item--item covariance matrix
[\citet{covsel}; \citet{friedman2008sparse}]---popularly
known as the
Graphical lasso.
This ensures an interpretable \textit{sparse} graph encoding of the
partial correlations,
and leads to favorable computational gains (as opposed to a dense
inverse covariance)
and also favors predictive performances. For the inverse covariance
regularization, we used
our C$++$ implementation of a primal block coordinate method applied to
the $\ell_1$ penalized
(negative) log-likelihood. Our algorithm [\citet{prox-glasso-2011}]
builds on \citet{friedman2008sparse} but has some important
differences, and scales better (for our experiments).
The main idea of the algorithm is outlined in Section~\ref
{M-step-glasso}, but we will
not elaborate on it since it is beyond the scope of this paper.

The rest of the paper is organized as follows. We describe our data
(with exploratory data analysis) in Section~\ref{secdata}. Modeling
details are provided in Section~\ref{secmodel}. This is
followed by model fitting details in Section~\ref{secfitting}. In
Section~\ref{secbireupg} we discuss the connection
between the widely-used bilinear random effects model (\textit{BIRE}) and
our proposed model. Section~\ref{secexp}
describes results of models fitted to Yahoo! front page data and
benchmark MovieLens data. We end with a
discussion in Section~\ref{secdisc}.

%s2 ###
\section{The PA module data}
\label{secdata}
Yahoo! front page (\href{http://www.yahoo.com}{www.yahoo.com}) is one
of the most visited content
pages on the web and receives hundreds of millions of user visits on
a daily basis. For a significant fraction of such visits, users
interact with some items on the PA module. We measure user interaction with
items through activities in a ``user session.''
A user session is a collection of visits by the user to the front page
where the inter-arrival time is less than 30 minutes [\citet
{cooley1999data}]. In each
such user session, if a recommended item is clicked, we interpret the
response to be positive
(i.e., labeled as 1). In the case of no click during the session, the
response is
negative (i.e., labeled as 0).

The illustrative front page data set used in this paper
contains around 5M binary observations generated by about 140K Yahoo!
users who interact with the PA module at least once over some period
spanning July to August 2009. A small random sample of sessions for
users who did not interact with the PA module at all during that time
period was also added. Although not using every user who visits
the front page may introduce bias in our parameter estimates,
the alternative approach of including all negatives introduced
significant noise and led to poor results. In fact, many users visiting
the front page do not interact with the PA module at all, hence,
negatives in our
data set lack perfect interpretation and are noisy. Such preprocessing
of negatives
to reduce noise is common in recommender problem studies reported in
the literature. For instance,
two widely used benchmark data sets, Netflix [\citet
{bennett2007netflix}] and
MovieLens [available at \href
{http://www.grouplens.org}{www.grouplens.org}], only include users
with more than $20$ ratings in
the training set. In most web application studies reported in the
literature, it is routine to
subsample negatives. We use the PA data described above to fit our
model and call it the
\textit{training} data. To test the accuracy of our models through
out-of-sample predictions, we created another \textit{test} data set
that contains observations from subsequent visits during
some time period in August 2009. To avoid bias in testing our methods,
the observations in the
test set are obtained through a randomized serving scheme---a small
fraction of randomly selected
visits are served with items that are randomly selected from the
available item set. Our randomized test set
contains approximately $528$K visits. Of these, about $300$K visits
were by users seen in the training set,
the remaining are by users who either did not visit during the training
period or were not included in the training set.

%f3 ###
\begin{figure}

\includegraphics{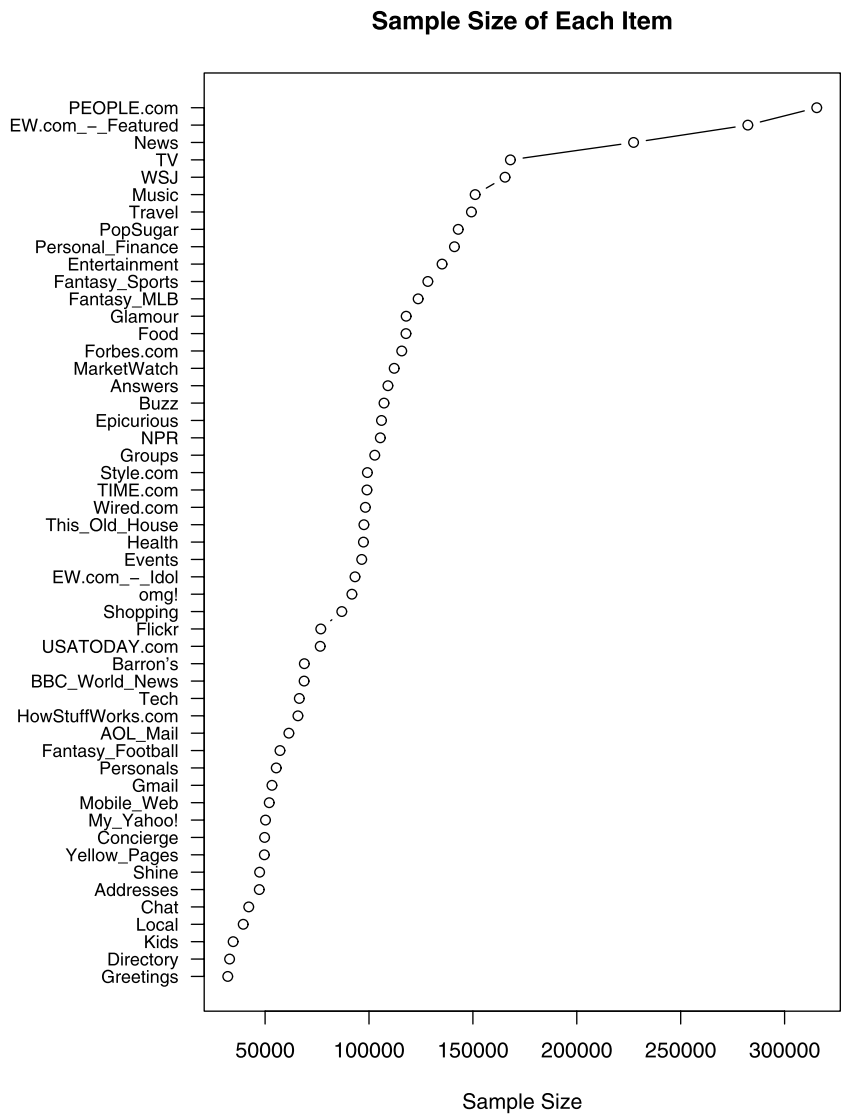}

\caption{The overall sample size of each item in the training data.}
\label{figsamplesizeCTR1}
\end{figure}

%f4 ###
\begin{figure}
\includegraphics{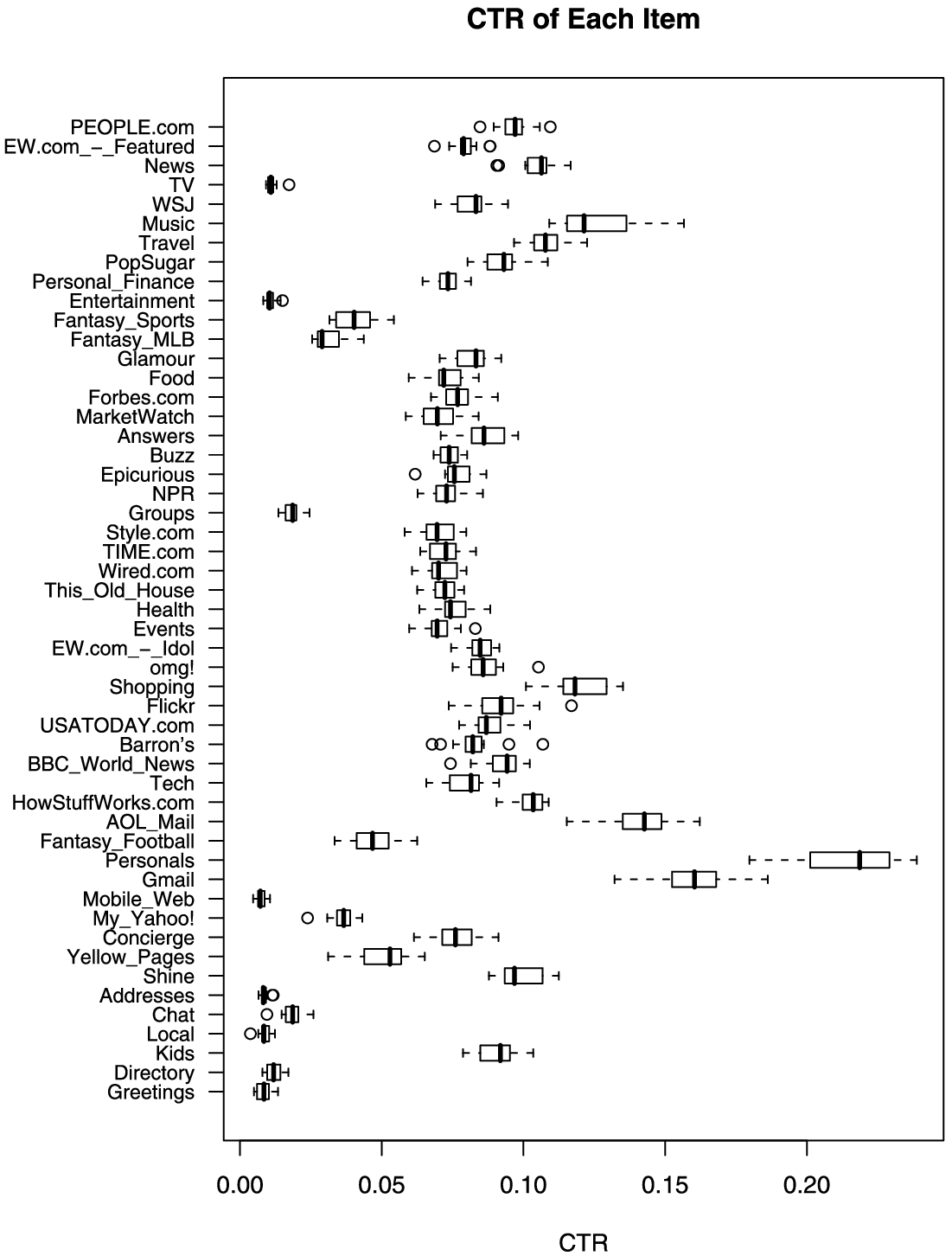}
\caption{The boxplot of the click through rate (CTR) per day for each item.}
\label{figsamplesizeCTR2}
\vspace*{5pt}
\end{figure}

We have a total of 51 items in our
data set, and on average each user viewed around 16 items during the
training period. The sample size and the click through rate (CTR) of
each item for the data sets used in this paper are shown in Figures~\ref{figsamplesizeCTR1} and
\ref{figsamplesizeCTR2}, respectively. For our
data, the CTR of an item is defined as the fraction of clicks per user
session. Clearly, CTR is in the range $[0,1]$.
As evident from Figure~\ref{figsamplesizeCTR2}, there is heterogeneity
in the CTRs of items; some items like
Personals,\footnote{\url{http://personals.yahoo.com/}.}
Gmail and
Music\footnote{\url{http://music.yahoo.com/}.} have relatively
high CTRs, while others like Mobile
Web,\footnote{\url{http://mobile.yahoo.com/}.}
Addresses\footnote{\url{http://address.yahoo.com/}.} and
Local\footnote{\url{http://local.yahoo.com/}.} have low
CTRs.\looseness=-1

A little more than half of the users in our data set were registered
when they visited Yahoo!; when these users visit the front page after
``logging in,'' we have
access to their demographic information like age, gender, occupation
and so
on. It is also possible to obtain a user's approximate geographic
location from the IP address. When a user is not logged in, we do not
have access to their demographic information. However, user activities
in both logged-in and logged-out states are tracked via the
browser cookie throughout the Yahoo! network; this helps us create a
browse signature for each cookie based on activity in a
set of categories like Sports, Autos, Finance and so
on [\citet{chen2009}]. The signature score in a given category is
based on user's visits to different Yahoo! websites, what ads they
clicked, what ads
they viewed, search queries issued by them and other activities. Thus,
for each user
(logged-in or logged-out), we have a few hundred covariates describing
their browse behavior
on the Yahoo! network. Each covariate is a binary indicator that is
turned on if a user is inferred
to have activity in the corresponding category.
For instance, if a user frequently visits
Yahoo! Music, the corresponding browse signature
covariate for Yahoo! Music will be 1; else it is 0.

%[Figure \ref{figdegree1}, Figure \ref{figdegree2} and Figure
%[Figure \ref{figdegree1}, Figure \ref{figdegree2} about here]

%f5 ###
\begin{figure}

\includegraphics{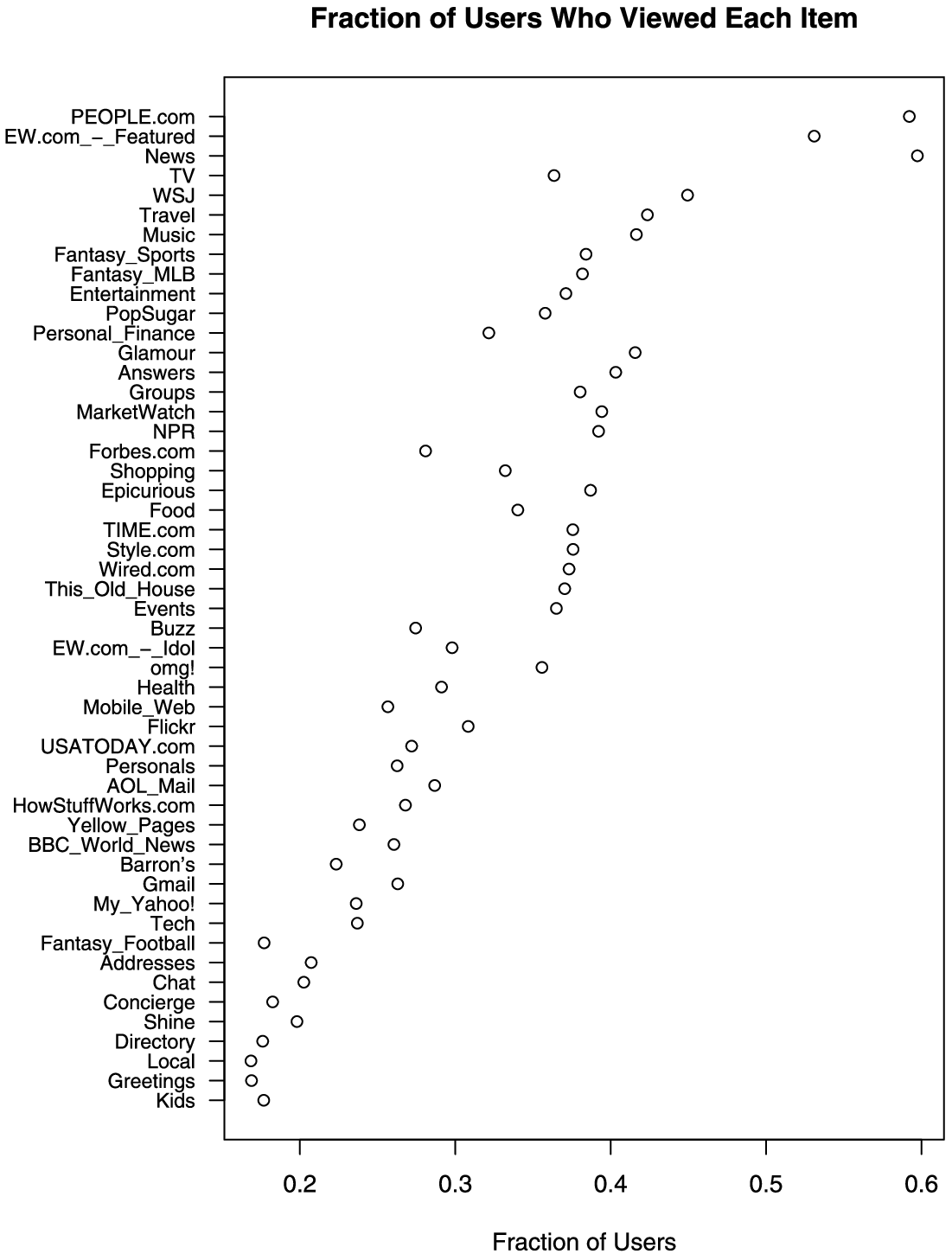}

\caption{Degree distribution for the items in the training set.}
\label{figdegree1}
\vspace*{5pt}
\end{figure}

%f6 ###
\begin{figure}

\includegraphics{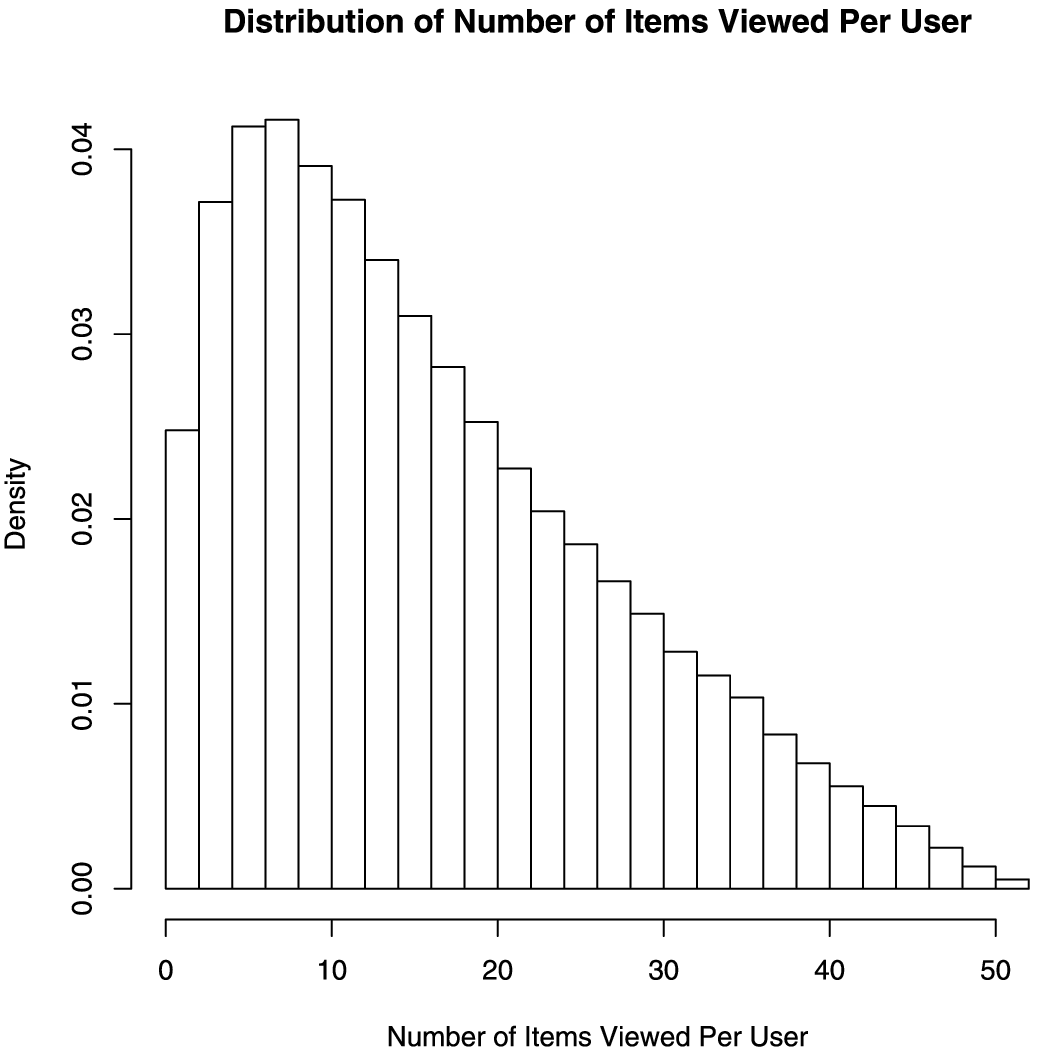}

\caption{Degree distribution for the users in the training set.}
\label{figdegree2}
\end{figure}

Figures \ref{figdegree1} and \ref{figdegree2} show the degree
distribution of items and users in the
training data. The distribution of the number of items clicked by users
has a peak at around~8, but the distribution is heavy tailed; a small
fraction of items are clicked by a large fraction of users. The degree
distribution clearly reveals the
imbalance in our data; modeling such data is challenging.

%s3 ###
\section{Detailed description of our models}
\label{secmodel}
We begin this section by describing the per-item logistic
regression model IReg introduced earlier. This is followed by our
per-user model that
assumes click rates on different items for a given user are
dependent. This dependence is modeled by associating a~random vector
per user and assuming the user random effects
are drawn from a~multivariate normal prior
with an unknown precision matrix. We impose sparsity on the precision matrix
through a Lasso penalty on the elements of the matrix. %
We shall call this the ``User Profile model with Graphical lasso''
(\textit{UPG}).

%s3.1 ###
\subsection{Per-item regression model: IReg}
Whenever applicable, we drop the time suffix from our notation.
For a user $u$ interacting with an item
$j$, we denote by $y_{uj}$ the binary response (click/no-click) in a session.
Then, $y_{uj}|p_{uj}\sim\operatorname{Bernoulli}(p_{uj})$ and $p_{uj}$
is modeled
through a logistic regression with log-odds denoted by $\theta_{uj}$:
%p_{uj} &=& \frac{1}{1+\exp(-\theta_{uj})} \\ \nonumber
%
%
%e1 ###
\begin{equation}
\label{logisticmodel}
p_{uj} = \frac{1}{1+\exp(-\theta_{uj})} \qquad \mbox{where } \theta
_{uj}=\mathbf{x}_u'\bolds{\beta}_j
\end{equation}
with $\mathbf{x}_u$ denoting the known per-user covariate vector which includes
age, gender and the user browse behavior information, and $\bolds{\beta
}_j$ is the
corresponding unknown item-specific regression coefficient vector
associated with item~$j$.

The user gender covariates include three categories: missing, male and~female.
Similarly, age is binned into 11 categories: missing, 0--12,
13--17, \mbox{18--20}, 21--24, 25--29, 30--34, 35--44, 45--54, 55--64, and older than
64. We also use 112 binary covariates describing user's browse behavior.
Finally, since we consider recommending items on two slots in region 2,
an extra categorical covariate was initially added to adjust for the
slot effect. In general, such adjustment for presentation bias is
important since everything else being equal, slots with less exposure
tend to have lower click-rates. We estimated the global slot effect by
looking at data obtained from a randomized
serving scheme. In our data set both slots provide similar exposure.

Since some items have a smaller number of observations in our data set,
the maximum likelihood
estimates of $\bolds{\beta}_j$ may not be stable for all $j$. Hence, we fit
our per-item regression models through a logistic ridge regression
fitted using the
library LIBLINEAR [\citet{fan2008liblinear}] that uses the trust
region Newton method [\citet{lin2008trust}]. Specifically, for each
application $j$, the (regualrized) coefficients are obtained by minimizing
%
%
%e2 ###
\begin{equation}
\frac{1}{2} \bolds{\beta}_j^{\prime}\bolds{\beta}_j +
C\sum_{u,j}\bigl(y_{uj}\log(p_{uj})+(1-y_{uj})\log(1-p_{uj})\bigr),
\end{equation}
where
%$y_{uj,rep}$ are replicates of the response
%variable $y_{uj}$ in the data and
$p_{uj}$ is given by equation~\ref{logisticmodel}. The tuning
parameter $C$ determines the amount of
shrinkage for the regression coefficients toward zero; we select it by
cross-validation.

\vspace*{3pt}
%s3.2 ###
\subsection{User profile model with gaphical lasso: {UPG}}
The \textit{IReg} model based on user covariates alone
fails to capture variability in item interactions per-user, especially
for heavy users with a large
number of previous visits. For
instance, there may be large variation in how users in a particular
age group interact with a Facebook item on PA.
For heavy users in the data, not accounting for such
variation could lead to an under-specified model. To ameliorate this,
we capture residual user-item interactions by augmenting the per-item
regression model with additional user-specific random
effects in the log-odds $\theta_{uj}$. More specifically, we introduce
random effects $\phi_{uj}$:
%
%
%e3 ###
\begin{equation}
\theta_{uj}=\mathbf{x}_u'\bolds{\beta}_j+\phi_{uj}.
\end{equation}
For user $u$, we denote the vector $\{\phi_{u1},\ldots,\phi_{uJ}\}$
as $\bolds{\phi}_u$, where $J$ is the total number
of items. This $J$ dimensional random vector captures user $u$'s
residual latent interaction with all $J$ items in the item set.
Obviously, this is an over-parameterized model, hence, the random
effects are constrained through
a prior distribution. We assume the following multivariate normal prior
distribution,
%
%
%e4 ###
\begin{equation}
\bolds{\phi}_u \sim\operatorname{MVN}(\mathbf{0},\bolds{\Sigma}),
\end{equation}
where $\bolds{\Sigma}$ is an unknown $J\times J$ covariance matrix and
$\bolds{\Omega} = \bolds{\Sigma}^{-1}$ is the
precision matrix.

In the training phase (model fitting) we obtain estimates
$\bolds{\hat{\beta}}_j$ for each\vspace*{-2pt} item~$j$ and the user preference
vector $\bolds{\hat{\phi}}_u$ for each user $u$. In the test period, if a
user $u$ has historical observations in training data so that
$\bolds{\hat{\phi}}_u$ is nonzero, then
%
%
%e5 ###
\begin{equation}
\hat{\theta}_{uj}=\mathbf{x}_u'\hat{\bolds{\beta}}_j+\hat{\phi}_{uj}.
\end{equation}
For a new user with no observations in the training period, we
fall back to the \textit{IReg} model and
%
%
%e6 ###
\begin{equation}
\hat{\theta}_{uj}=\mathbf{x}_u'\hat{\bolds{\beta}}_j.
\end{equation}
Note that for $\bolds{\hat{\phi}}_u$ to be nonzero, it is enough to
have partial response information on a
subset of $J$ items for user $u$. The random effects corresponding to
the missing response items
for user $u$ are estimated by combining the likelihood of observed user
response with the global prior
on user random vector. The global prior that is completely specified by
the precision matrix is estimated by
pooling data across all users.

Although the $J\times J$ precision matrix $\bolds{\Omega}$ provides an estimate
of pairwise similarities between items after adjusting for the
effects of user covariates and other items, the total number of
parameters to
estimate for $J$ items is large ($O(J^2)$). Such estimation may get difficult
for large $J$. It is thus desirable to impose further regularization on
$\bolds{\Omega}$, to avoid overfitting and improve the predictive performance.
%may be required to get estimates with lower-variance.
%adding
%further constraints on the elements of the precision matrix may help
%improve performance.
%This is popularly dubbed as a inverse-covariance regularization
%With i.i.d observations from a multivariate normal distribution, this
%is also known as the
%It is well known that covariance
%selection is difficult for high dimensional problems.
High-dimensional (inverse) covariance estimation is a challenging
problem; see \citet{covsel}, \citet{dempster1972},
\citet{esl-09},
\citet{lauritzen1996graphical} and
references therein. The form of regularization depends upon the nature
of the problem, dimension of the parameter-space and computational
considerations among others.
A diagonal covariance, for example,
is practically unrealistic since items are far from marginally independent.
Regularization schemes resulting in dense and possibly unstructured precision
graphs lack interpretability and will lead to increased computational burden.
We propose using a regularization scheme that encourages sparsity in
the precision matrix,
popularly referred to as the\vadjust{\goodbreak} sparse-inverse covariance selection or
Graphical Lasso [\citet{covsel}; \citet{friedman2008sparse}].
%encodes a graph of
%conditional dependencies among items, since items are likely to be
%conditionally independent
%a sparse covariance structure
%in most recommender problems as several items tend to be
%similar and user interactions are expected to be correlated.
%covariance matrix is a possible choice but this assumes there is no
%residual dependence in the click-rates of items. Intuitively, this is
%unlikely to be the case in most recommender problems as several items
%tend to be
%similar and user interactions are expected to be
%correlated.
%However, it is expected that the correlations would exist
%only for a small fraction of items pairs, hence a sparse
%structure for $\bolds{\Omega}$ is more reasonable, especially for large
%$J$.
%Other than improving accuracy, sparseness also helps in expediting
%model fitting as we shall see later.

Introducing sparsity into the precision matrix is a well-studied
problem, especially in the context of graphical models with Gaussian
data [\citet{lauritzen1996graphical}]. In fact, for Gaussian models
$\bolds{\Omega}_{r,s} = 0$ implies $\phi_{ur}$ and $\phi_{us}$ are
conditionally independent given the rest of the coordinates.
Thus,
sparsity in the precision matrix learns the structure of the graphical
model.
Due to the hierarchical model structure proposed, the estimated
covariance/precision matrices have to be positive definite, rendering
multiple regression or pseudo-likelihood
based approaches like \citet{besag-75}, \citet{MB2006}
unsuitable for our task.
The Graphical Lasso (Glasso) method estimates the precision matrix by
minimizing the following regularized negative log-likelihood criterion:
%proposed by \cite{friedman2008sparse}.
%Glasso ensures
%sparsity in the precision matrix by imposing a lasso penalty on the
%elements of the precision matrix.
%style algorithm pursued,
%The challenge is to impose both
%sparsity and yet obtain an estimate that is positive definite. In
%other words, the support of our parameter space is the set of
%precision matrices $\bolds{\Omega}$ that are positive definite but
%satisfy the $L_1$ constraint $\norm{\bolds{\Omega}}_1 <
%where the constant
%controls the amount of shrinkage that is applied to the entries of
%$\bolds{\Omega}$. In fact, if $\mathbf{S}$ is an empirical covariance
%matrix
%of the random effects $\{\bolds{\phi}_{u}\}$, then following
%by
%minimizing
%
%
%e7 ###
\begin{equation}
\label{glassoproblem}
-\log\det{\bolds{\Omega}}+\operatorname{tr}(\mathbf{S}\bolds{\Omega
})+\rho\|
\bolds{\Omega}\|_1 \qquad \mbox{with } \bolds{\Omega} \succ0.
\end{equation}
%
%subject to $\bolds{\Omega} \succ0$.
Here, the quantity $\| \bolds{\Omega}\|_1$ denotes the sum of absolute
values of the
matrix $\bolds{\Omega}$.
The parameter $\rho$ controls
the amount of $L_1$ regularization and the sparsity induced on the
estimated precision matrix. The optimization problem in equation (\ref
{glassoproblem}) above is convex [\citet{BoydConvex}].
%and it is implemented in Glasso by iteratively solving a set of $J$
%standard Lasso
%regressions \citep{tibshirani1996regression}. Each Lasso is solved
%through a coordinate
%descent algorithm \citep{friedman10glmnet}.
In our model fitting
procedure, the sample covariance matrix $\mathbf{S}$ is obtained in the
E-step by taking expectation of the log-prior with respect to the
posterior distribution of $\bolds{\phi}_u$'s assuming the hyper-parameter
$\bolds{\Omega}$ is fixed at the latest estimate in the EM procedure (see
Section~\ref{secfitting} for complete details).

To reiterate, the \textit{UPG} model offers the following advantages
over \textit{IReg}:
%a). It accounts for variability in click behavior at user
%level, this is In our data it is obvious to see that some users
%like to click more than some others. The mean
%$\sum_{i}\phi_{ui}/I$ addresses this point.
%
\begin{itemize}
\item\textit{UPG} accounts for residual variation in user preference
for items
after adjusting for covariates through \textit{IReg}. Since user covariates
include only coarse behavioral attributes inferred
based on user activity across the Yahoo!
network, it may not completely reflect user preferences for PA
items.
\item\textit{UPG} exploits item--item similarity to infer user
preference on items that he/she may have not been exposed
to before, for instance, if users who click on item X also
tend to click on Y in the historic data; a click by a~new user on X
would imply a click on Y with a high probability.
\end{itemize}

%s4 ###
\section{Model fitting procedure}
\label{secfitting}
The model fitting procedure for the \textit{IReg} model
is implemented through a trust region Newton
method as described in Section~\ref{secmodel}. In this
section we describe the fitting procedures for our UPG model.
We fit our UPG model through a penalized quasi-likelihood (PQL) method
[\citet{breslowclayton}]. Although other fitting methods based on MCMC
and better approximation of the marginal likelihood through
Gauss--Hermite quadrature [\citet{bates}] are possible, we use PQL for
scalability. In fact, preliminary experiments conducted using MCMC
methods clearly revealed the difficulty of scaling to data sets
analyzed in this paper. In particular, we tested the MCMC method in the
statistical software R based on Langevin Metropolis--Hastings
algorithms [\citet{roberts2001optimal}] using a diagonal precision
matrix as prior (random walk Metropolis proposal was slow to converge).
Even for this simplified model, we need approximately 7K
posterior draws of $\hat{\bolds{\phi}}_u$ per user to obtain $1$K
posterior samples (based on MCMC diagnostics, a burn-in of $2$K and
thinning of $5$ was
adequate). Obtaining samples for all users in our training data took
approximately $7$ days for the PA data due to high dimensionality of
random effects and a large number of users.
More crucially, a sampling scheme with $L_1$ penalty on the elements of the
precision matrix is nontrivial to construct since it is not clear
what prior on the precision matrix would be equivalent to
$L_1$ regularization. For regression problems, $L_1$ penalty is
equivalent to a double exponential prior on the regression
coefficients, but this does not hold in our case due to the additional
constraint of positive definiteness. Hence, we take advantage of the
Glasso mechanism for estimating $\bolds{\Omega}$ and to do
so, we found the PQL procedure more amenable.
Thus, in this paper we
only focus on PQL and leave the exploration of other fitting
procedures to future work. For instance, a parametric bootstrap
procedure discussed in \citet{Kuk95} can perhaps be modified
to remove any bias incurred in estimating the precision matrix.
Bootstrap is a better strategy for large scale application like ours since
multiple runs can be performed in parallel.\looseness=-1

Before describing the PQL fitting procedure, we begin with some notation.
Let $\bolds{\beta}=\{\bolds{\beta}_1,\ldots,\bolds{\beta}_J\}$ be the
set of
regression coefficients,
and $\bolds{\Theta} = (\bolds{\beta},\bolds{\Omega})$ denote the fixed
effects to be estimated in the UPG model.
The PQL method works as follows---at the current value $\bolds{\Theta
}=\bolds{\Theta}_{0}$, we form ``working residuals''~$Z_{uj}$ corresponding
to the response $y_{uj}$ (the response for user
$u$ and item~$j$) through a Taylor series expansion. The residuals are
used to obtain the posterior distribution of random-effects at $\bolds
{\Theta}_{0}$ (E-step). This is followed
by an updated estimate of $\bolds{\Theta}$ in the M-step. The formation
of working residuals and the EM steps
on the working residuals are iterated until convergence. We note that
for Gaussian responses, working residuals
coincide with true responses and no approximation is incurred. The
complete mathematical details on the PQL procedure for fitting UPG are
provided below.

%s4.1 ###
\subsection{Algorithm for learning the \textit{UPG} model} \label
{seclearn-upg}
If $\hat{\phi}_{uj}$
and $\hat{\bolds{\Theta}} = (\hat{\bolds{\beta}},\hat{\bolds{\Omega}})$
denote the current
estimates of the random-effects and parameters, respectively,
the working residual $Z_{uj}$ for binary response $y_{uj}$ is given by
%
%
%e8 ###
\begin{equation}
Z_{uj} = \hat{\eta}_{uj} + \frac{y_{uj} - \hat{p}_{uj}}{\hat
{p}_{uj}(1-\hat{p}_{uj})},
\end{equation}
where
\[
\hat{\eta}_{uj} = \mathbf{x}_{u}^{\prime}\hat{\beta}_{j} + \hat{\phi}
_{uj} \quad \mbox{and} \quad
\hat{p}_{uj} = \frac{1}{1+\exp(-\hat{\eta}_{uj})}.
\]
With these we have approximately,
%
%
%e9 ###
\begin{equation}\label{laplace}
Z_{uj}\sim N(\phi_{uj}+\mathbf{x}_{u}'\bolds{\beta_j},V_{uj}) \qquad
\mbox
{where }V_{uj} = \bigl(\hat{p}_{uj}(1-\hat{p}_{uj})\bigr)^{-1}.
\end{equation}
The updated estimates of random effects and parameters are
now obtained by solving the model in equation (\ref{laplace}) through an
EM algorithm [see \citet{breslowclayton} for more details on PQL
in general].
The EM algorithm treats the random effects as missing data [\citet
{dempster1977maximum}].
Thus, the E-step involves computing the expected log-likelihood of the
complete data with respect to the conditional distribution of random effects
given $\hat{\bolds{\Theta}}, \mathbf{Z}$.

Let $e_{uj}=Z_{uj}-\mathbf{x}_u'\hat{\bolds{\beta}}_{j}$. Denote by
$n_{uj}$ the number of replicates where user~$u$ interacts with item $j$,
and let $N_u$ be the total number of users.
Also, let
\[
\mathbf{K}_u = \operatorname{diag}\biggl(\frac{n_{u1}}{V_{u1}},\ldots,\frac
{n_{uJ}}{V_{uJ}}\biggr), \qquad
\mathbf{U}_u =
\Biggl(\sum_{r=1}^{n_{u1}}\frac{e_{u1,r}}{V_{u1}},\ldots,\sum
_{r=1}^{n_{uJ}}\frac{e_{uJ,r}}{V_{uJ}}\Biggr),
\]
where $e_{uj,r}$ represents the $r$th replicate for $e_{uj}$.\vadjust{\eject}

The conditional distribution of $\bolds{\phi}_u$ given $\hat{\bolds
{\Theta}},\mathbf{Z}$ is
%
%
%e10 ###
\begin{equation}\label
{eqnphiposterior}
\bolds{\phi}_u|\mathbf{Z},\hat{\bolds{\Theta}} \sim\operatorname
{MVN}(\bolds
{\mu}_{u},\bolds{\Sigma}_{u}),
\end{equation}
where,
%
%e11 ###
\begin{equation}
\label{eqnphiposterior1}
\bolds{\mu}_{u}=(\mathbf{K}_u+\hat{\bolds{\Omega}})^{-1}\mathbf{U}_u,
\bolds
{\Sigma}_{u} = (\mathbf{K}_u+\hat{\bolds{\Omega}})^{-1},
\end{equation}
where $\bolds{\mu}_{u}=(\mathbf{K}_u+\hat{\bolds{\Omega}})^{-1}\mathbf{U}_u$,
and $\bolds{\Sigma}_{u} = (\mathbf{K}_u+\hat{\bolds{\Omega}})^{-1}$. Thus,
$\hat{\bolds{\phi}}_u = \bolds{\mu}_{u}$ and the updated values of
$\bolds
{\Omega}$ and $\bolds{\beta}$ are obtained as follows.
We use $\hat{\bolds{\phi}}_{u}$ as offset and update the estimates of
$\bolds{\beta}$ through the LIBLINEAR routine. To update $\bolds{\Omega
}$, we make use of the following observation:
%
%
%e12 ###
\begin{eqnarray}
E_{\bolds{\phi}|\hat{\bolds{\Omega}},\mathbf{Z}}\biggl[\sum_{u}\log p(\bolds
{\phi
}_u|\bolds{\Omega})\biggr] &=& -\frac{pN_u}{2}\log(2\pi) + \frac{N_u}{2}{\log}
|\bolds{\Omega}|\nonumber
\\[-8pt]
\\[-8pt]
&&{}-\frac{1}{2}\sum_u \operatorname{tr}(\bolds{\Omega
}\bolds
{\Sigma}_u)+\bolds{\mu}_u'\bolds{\Omega}\bolds{\mu}_u.
\nonumber
\end{eqnarray}
The updated value of the precision matrix, that is, $\bolds{\Omega}$, is
obtained from the regularized likelihood criterion
(\ref{glassoproblem}). In particular, for the special case with $\rho
=0$, corresponding to the unregularized maximum
likelihood, the covariance estimate (assuming it exists) is given by
%The updated value of $\bolds{\Omega}$ without the Lasso penalty in the
%M-step is then obtained as
%
%
%e13 ###
\begin{equation}\label{eqnMstep}
\hat{\bolds{\Omega}}^{-1} = \frac{\sum_u(\bolds{\Sigma}_u+\bolds{\mu
}_u\bolds{\mu}_u')}{N_u}.
\end{equation}

When $\rho\neq0$, we treat the result of equation (\ref{eqnMstep})
as the sample covariance matrix $\mathbf{S}$, and use the Glasso
regularization to obtain a sparse $\hat{\bolds{\Omega}}$ and
corresponding covariance matrix $\hat{\bolds{\Sigma}}$.

%s4.1.1 ###
\subsubsection{Large scale implementation of the E-step}
\label{secupgcomp}
With $N_u$ users and~$J$ items, a less sophisticated implementation of
the E-step (to obtain $\bolds{\Sigma}_u$ and~$\bolds{\mu}_u$ for all
the users)
is at least $O(N_uJ^3)$, due to the expensive
computation for computing $\bolds{\Sigma}_u = (\mathbf{K}_u+\hat{\bolds
{\Omega}})^{-1}$ (\ref{eqnphiposterior1}). This can
make the training process prohibitively slow when $J$ is large (e.g., a few
thousands). However, we note that the matrix inversion need not be done
from scratch.
If $\hat{\bolds{\Omega}}^{-1}$ is available from the previous iteration
of the EM algorithm,\footnote{While using the
Glasso regularization in the M-step, we see that this is\vspace*{-2pt} indeed the
case, since the algorithm returns both $\hat{\bolds{\Omega}}^{-1}$
and $\hat{\bolds{\Omega}}$, without the cost of an explicit inversion.}
$(\mathbf{K}_u+\hat{\bolds{\Omega}})^{-1}$ can be obtained via
a low-rank update, where the low-rank is given by the number of nonzero
entries of the diagonal matrix $\mathbf{K}_u$. Since in most recommender
problems, a large fraction of users only
interact with a small fraction of items,
$\| \mathbf{K}_u\|_0$---the number of nonzero diagonal values in
$\mathbf{K}_u$, is usually small.
%To see how this is achieved, we will first assume, WLOG that the items
%are ordered such that
%the nonzero diagonal entries of $\mathbf{K}_u$ are located in the top $
An application of the Sherman--Morrison--Woodbury formula gives
%
%
%e14 ###
\begin{equation} \label{sherman-1}
(\hat{\bolds{\Sigma}}^{-1}+\mathbf{K}_u)^{-1} =
\hat{\bolds{\Sigma}} -\hat{\bolds{\Sigma}} \sqrt{\mathbf{K}_u}\bigl(\mathbf
{I}+\sqrt{\mathbf{K}_u}\hat{\bolds{\Sigma}}\sqrt{\mathbf
{K}_u}\,\bigr)^{-1}\sqrt
{\mathbf{K}_u}\hat{\bolds{\Sigma}},
\end{equation}
where computing
$(\mathbf{I}+\sqrt{\mathbf{K}_u}\hat{\bolds{\Sigma}}\sqrt{\mathbf{K}_u})^{-1}$
%(note that this is a $\norm{\mathbf{K}_u}_0 \times\norm{
%matrix)
takes only $O(\| \mathbf{K}_u\|_0^3)$.
%Hence the covariance $\Sigma_u$ is obtained in $O(J^2)$ operations, an
%order of magnitude improvement over
%$O(J^3)$!.
However, we show below that this can be even more efficient by
exploiting the structure of the sufficient statistics.

Note that in the M-step the object of interest is actually
$\sum_u \bolds{\Sigma}_u$ instead of the individual $\bolds{\Sigma
}_u$'s. Using equation (\ref{sherman-1}), we have the following simplification:
%
%
%e15 ###
\begin{eqnarray}\label{sherman-2}
\sum_u \bolds{\Sigma}_u & = & \sum_u (\hat{\bolds{\Sigma}}^{-1}+\mathbf
{K}_u)^{-1}\\ \nonumber
& = & N_u\hat{\bolds{\Sigma}} -\hat{\bolds{\Sigma}} \biggl(\sum_u
\sqrt{\mathbf
{K}_u}\bigl(\mathbf{I}+\sqrt{\mathbf{K}_u}\hat{\bolds{\Sigma}}\sqrt{\mathbf
{K}_u}\bigr)^{-1}\sqrt{\mathbf{K}_u}\biggr)\hat{\bolds{\Sigma}}.
\end{eqnarray}
The computational cost in obtaining $\sum_u\!\bolds{\Sigma}_u$ using
(\ref{sherman-2}) is now
$O(\sum_u \!\| \mathbf{K}_u\|_0^3)+O(J^2)$.
Alternatively, using decomposition (\ref{sherman-1}) and then summing
over all users to obtain
$\sum_u \bolds{\Sigma}_u$ has a
complexity of $O(\sum_u \| \mathbf{K}_u\|_0^3 ) + O(N_uJ^2)$, which is
definitely much higher than
$O(\sum_u \| \mathbf{K}_u\|_0^3) + O(J^2)$ for large values of~$N_u$.
We would like to point out that a sparse precision matrix $\hat{\bolds
{\Omega}}$ does not lead to much computational benefit in the
strategies just described via (\ref{sherman-1}) and~(\ref
{sherman-2}). This is because we
operate on covariance matrices which may still be dense even if the
corresponding precision matrices are sparse. The sparsity of the
precision matrices, however, plays a crucial role in updating the
posterior mean $\bolds{\mu}_u$ for each user $u$.
% hence the sparsity in
%$\bolds{\Sigma}^{-1}$ does not play much of a role not been exploited
%$\sum_u \sqrt{\mathbf{K}_u}(\mathbf{I}+\sqrt{\mathbf{K}_u}\hat{\bolds{
%$(\mathbf{I}+\sqrt{\mathbf{K}_u}\hat{\bolds{\Sigma}}\sqrt{
%takes $O(\norm{\mathbf{K}_u}_0^3)$.

For obtaining $\bolds{\mu}_u$, we need to solve the sparse symmetric linear
systems $(\mathbf{K}_u+\hat{\bolds{\Omega}})\bolds{\mu}_u = \mathbf
{U}_u$ for
all the users.
%we use the
%conjugate gradient method to solve the
%linear system $(\mathbf{K}_u+\hat{\bolds{\Omega}})\bolds{\mu}_u =
Direct factorization based methods can be quite expensive for
solving these linear systems for arbitrary sparsity patterns in $\hat
{\bolds{\Omega}}$.
For this purpose we employ iterative methods based on conjugate gradients
[\citet{demmel-cg}; \citet{hestenes1952methods}]. For a
specific user, this method returns approximate solutions to the
linear system at the cost of a few multiplications of the matrix
$(\mathbf{K}_u+\hat{\bolds{\Omega}})$ with a vector, hence
the complexity being \textit{linear} (or better) in $J$ for
sufficiently sparse $\hat{\bolds{\Omega}}$. For dense~$\bolds{\Omega}$,
however, the computational cost increases to $O(J^2)$
(see Section~\ref{sectimes} for computational results).
Though well-chosen pre-conditioners can decrease
the number of conjugate gradient iterations, all the experimental
results reported in this paper are without any pre-conditioning.

Finally, since the computation of $\sum_u\bolds{\Sigma}_u$ and $\sum
_u\bolds{\mu}_u\bolds{\mu}_u'$ can easily be parallelized across users,
we have used multiple-threading (e.g., 7 threads) to further expedite
the computations.
%we use multiple threads (e.g. 7) to expedite
%the computation of E-step in UPG.
%s4.1.2 ###
\subsubsection{Computational considerations in the M-step: The $l_1$
regularized log-likelihood} \label{M-step-glasso}

For the $l_1$ regularized log-likelihood, that is, Glasso
regularization in the M-step, the input
covariance matrix is
%$\mathbf{S} =\frac{\sum_u(\bolds{\Sigma}_u+\bolds{\mu}_u\bolds{
$\mathbf{S} =\sum_u(\bolds{\Sigma}_u+\bolds{\mu}_u\bolds{\mu}_u')/ N_u.$
The Glasso implementation [\citet{friedman2008sparse}] involves
computation of $J$ Lasso regressions, and a $J
\times J$ Lasso regression has a worst case complexity $O(J^3)$, hence,
the computational
complexity of Glasso could be as high as $O(J^4)$ in the worst
case. However, significant computational advances in Lasso type
computations can make this computation faster, especially for large
sparsity parameter $\rho$. Many such computational nuances have
already been incorporated into the Glasso code by \citet
{friedman2008sparse}. In particular, each
Lasso is performed through a fast coordinate descent procedure that yields
computational savings through residual and active set
updates. In fact, in \citet{friedman2008sparse} the authors empirically
demonstrate
that computational complexity of Glasso is $O(J^3)$ for dense problems
and much
faster for sparse problems with large $\rho$.
Our algorithm [see \citet{prox-glasso-2011} for details] enjoys
the the major
computational advantages of the Glasso algorithm of \citet
{friedman2008sparse}.
As mentioned earlier, we
choose to avoid the details of our algorithm, since it is beyond the
scope of this current paper. We outline some of its salient features,
leaving the details to a future paper.
It is essentially a primal block-coordinate method, which
also requires solving for every row/column a
Lasso problem. In fact, a partial optimization in the Lasso problems
suffices, for convergence to hold.
Our algorithm maintains \textit{both} the precision and its inverse,
that is, the
covariance matrix along the course of the algorithm, and is amenable to
early stopping.
This is actually a crucial advantage of our method, which, as our
experiments suggest, it struggles with the implementation of
\citet{friedman2008sparse}.
Our algorithm has
very competitive computational complexity as the Glasso for sparse
problems, as our experiments suggest.
It goes without saying that the algorithm of
\citet{friedman2008sparse}, when compared with ours, gives the same
models, upon convergence---since they both
solve the same convex optimization criterion.
However, we chose our algorithm for our experiments
since it scaled favorably for our experiments, especially for the
MovieLens data with approximately $4000$ items.

Based on the above discussion, we conclude that large values of $\rho$
that induce
more sparsity in $\bolds{\Omega}$ have two computational advantages: it
speeds up the conjugate gradient
computations in the E-step and it also accelerates the M-step of
our algorithm.

%We also note that a sparse $\bolds{\Omega}$ expedites the computation
%of $
%conjugate gradient step to obtain $\b
%m{\mu}_u$.
%}
%We also note that a sparse $\bolds{\Omega}$ expedites the computation
%of
%posterior in %Equation~(\ref{eqnphiposterior}) through a sparse
%Cholesky decomposition routine \citep{chen2006algorithm}.

%Since Glasso is remarkably fast and can be treated as a black-box
%algorithm, we insert Glasso to the user profile model fitting as
%follows: For each iteration, we update the conditional posterior
%$p(\bolds{\phi}_u|\mathbf{Y}_u,\hat{\bolds{\beta}},\hat{\bolds{
%sparse matrix inversion library CHOLMOD \citep{chen2006algorithm}, and
%then we update $\hat{\bolds{\Sigma}}$ by EM algorithm using equation
%(\ref{Mstep}). Before we go the next iteration, we use Glasso by
%treating the sample covariance matrix as $\hat{\bolds{\Sigma}}$ and
%thus
%obtain the corresponding
%$\bolds{\Theta}$ and $\bolds{\Theta}^{-1}$. $\hat{\bolds{
%updated to be the sparse matrix $\bolds{\Theta}^{-1}$ for the next
%iteration.

%s4.2 ###
\subsection{The \textit{UPG}-online model}
We have observed that
updating the posterior distribution of $\bolds{\phi}_u$'s in an online
fashion, as new observations are obtained, leads to
better predictive performances.
This is because for a large fraction of users, the posterior of $\bolds
{\phi}_u$ is based on a small
number of visits in the training period. Keeping per-user posteriors updated
using all prior user visits leads to
posterior estimates that provide a better model fit. We observed that
it is not necessary to update the precision matrix $\bolds{\Omega}$ too
quickly if the item set does
not change. This is because $\bolds{\Omega}$ is a global parameter
estimated by using large amounts of training data.
We now describe how the posterior of $\bolds{\phi}_u$ gets updated
online with the arrival of a new observation~$y_{uj,\mathrm{new}}$ for
a fixed $\bolds{\Omega}$.

Let the current posterior of $\bolds{\phi_{u}} \sim \operatorname{MVN}(\bolds{\mu
}_{u},\bolds{\Sigma}_{u})$ as given in equation~(\ref{eqnphiposterior}).
Assuming a new response $y_{uj,\mathrm{new}}$ updates the counters $\mathbf{K}_u$
and $\mathbf{U}_u$ to $\mathbf{K}_{u,\mathrm{new}}$ and $\mathbf{U}_{u,\mathrm{new}}$,
the new
posterior mean
$\bolds{\mu}_{u,\mathrm{new}}$ is given by solving $(\mathbf{K}_{u,\mathrm{new}}+\hat
{\bolds
{\Omega}})\bolds{\mu}_{u,\mathrm{new}}=\mathbf{U}_{u,\mathrm{new}}$ through conjugate
gradient as
described in Section \ref{secupgcomp}. We note that for a sparse $\bolds
{\Omega}$, this computation is fast even for a large $J.$ Also note that
posterior variance need not be updated explicitly, as it is
automatically updated implicitly once we obtain an updated $\mathbf{K}_u.$
This is
an important implementation detail for large recommender problems; we
only need to store the counters in $\mathbf{K}_u$ and $\mathbf{U}_u$ for
items that have
been shown to users in the past. This reduces the memory requirements
per user and helps with scalability in large scale systems with
hundreds of
millions of users. The actual implementation that scales to large
number of users that visit Yahoo! front page requires state-of-the-art
databases like
BigTable and Hbase that can store petabytes of data across multiple
commodity servers and among other things, they can support realtime
serving. While it is difficult to produce the result of such an
experiment for an academic paper,
such implementations are becoming more common for large web
applications. They may, however, involve significant hardware and other
engineering costs
that are typically affordable for recommender applications with a
massive number of visits.

%s5 ###
\section{Comparing \textit{BIRE} and \textit{UPG}}
\label{secbireupg}
In this section we
discuss the connection between \textit{BIRE} and \textit{UPG} since the
former is considered state-of-the-art in the existing recommender
literature. To simplify the discussion,
we will assume that our response $y_{uj}$ is Gaussian. We also assume
our response has been adjusted to remove the effects of covariates and
main effects. Hence,
the \textit{UPG} model in this case is given as
%
%
%e16 ###
\begin{equation}
\label{upgguassian}
y_{uj}|\phi_{uj} \sim N(\phi_{uj},\sigma^2) \qquad
\mbox{where } \bolds{\phi}_{u} \stackrel{\mathrm{i.i.d.}}{\sim}
\operatorname{MVN}(\mathbf{0},\bolds{\Omega}^{-1}).
\end{equation}
Conditional on the $\bolds{\phi_u}$'s, the responses are independent but
marginally they are not. In fact, responses by the same user $u$ on
different items
are dependent. Denoting by $\mathbf{Y}_u$ the response of user $u$ on $J$
items, we see
\[
\mathbf{Y}_u \sim \operatorname{MVN}(\mathbf{0},\sigma^2\mathbf{I} + \bolds{\Omega}^{-1})
\]
and the $L_1$ regularization on the precision, as we have described
before helps both in computation and predictive performance.
%especially for large $J.$

For the \textit{BIRE} model, we have
%
%
%e17 ###
\begin{eqnarray}
\label{upgguassian}
y_{uj}|\mathbf{q}_{u},\mathbf{v}_j \sim N(\mathbf{q}_{u}^{\prime}\mathbf{v}_j,\sigma^2)\nonumber
\\[-8pt]
\\[-8pt]
\eqntext{\mbox{where }\mathbf{q}_u \stackrel{\mathrm{i.i.d.}}{\sim} \operatorname
{MVN}(\mathbf{0},I)
 \mbox{ and }
\mathbf{v}_j \stackrel{\mathrm{i.i.d.}}{\sim} \operatorname{MVN}(\mathbf{0},aI).}
\end{eqnarray}
Here $\mathbf{q}_u$ and $\mathbf{v}_j$ are $K$-dimensional user and item
random effects (also called factors).
Marginalizing over user factors $\mathbf{q}_u$, we see
\[
\mathbf{Y}_u \sim \operatorname{MVN}(\mathbf{0},\sigma^2\mathbf{I} + \mathbf{V}\mathbf{V}^{\prime}),
\]
where $\mathbf{V}$ is a $J \times K$ matrix of item factors stacked
together. Thus, the \textit{BIRE} model
estimates item--item similarities through a low-rank decomposition in
contrast to the UPG model that assumes a more
general structure, with the sparsity regularization controlling the
degrees of freedom of the estimator.
% can be regulrized vua regularized through sparsity
%inducing $L_1$ prior.
%s5.1 ###
\subsection{Computational complexity of fitting BIRE}
\label{seccompcomplexbire}
Several model fitting strategies have been used to fit the \textit{BIRE}
model in the literature. Of these, stochastic gradient descent (SGD) and
Monte Carlo EM (MCEM) have emerged as methods of choice. We note that
since the posterior is multi-modal, model fitting methods
influence the local optima obtained that in turn affects prediction
accuracy. In particular, MCEM seems to provide the best performance in
terms of out-of-sample
predictive accuracy [\citet
{rlfm}; \citet{salakhutdinov2008bayesian}]. The
E-step of the MCEM procedure computes the expected log-posterior by
drawing samples from the
posterior of $\{\mathbf{q}_u,\mathbf{v}_j\dvtx  u=1,\ldots,N_u;j=1,\ldots,J\}$,
conditional on the current hyper-parameter estimate $a$ (and $\sigma
^2$ for Guassian responses). We note that conditional on $\mathbf{V}$, the
$J \times K$ matrix of item factors stacked together, the~$\mathbf{q}_u$'s are independent.
Similarly, conditioning on~$\mathbf{Q}$, the $N_u \times K$ matrix of user factors,
$\mathbf{v}_j$'s are independent. This provides a simple Gibbs sampling
strategy to obtain samples from the posterior. The posterior samples
are used to obtain an estimate
of the expected log-prior with respect to the latest posterior
which is then used to obtain an updated estimate of $a$ (and $\sigma
^2$ when applicable) in the M-step. In fact, it is trivial to see that
the updated estimate of $a$ (and
$\sigma^2$ when applicable) in the M-step is obtained in closed form;
thus, the computational complexity of the fitting procedure is mainly
due to the E-step.
For Guassian responses,\vadjust{\goodbreak} the conditional distribution of $(\mathbf
{q}_u|\mathbf
{V},a,\sigma^2)$ is Guassian with
mean and variance given by
\begin{eqnarray*}
\operatorname{Var}(\mathbf{q}_u|\mathrm{Rest}) &=& \biggl(I + \sum_{j \in
\mathcal
{N}_u} \frac{\mathbf{v}_j \mathbf{v}_j^{\prime}}{\sigma^{2}} \biggr)^{-1}, \\
E(\mathbf{q}_u|\mathrm{Rest}) &=&\operatorname{Var}(\mathbf{q}_u|\mathrm
{Rest})\sum
_{j \in\mathcal{N}_u}\frac{y_{uj}\mathbf{v}_j}{\sigma^2},
\end{eqnarray*}
where $\mathcal{N}_u$ denotes the set of items user $u$ interacted
with, and $\| K_{u}\|_0$ denotes the size of this set.
The computational complexity of computing the outer-product in
$\operatorname
{Var}(\mathbf{q}_u|\mathrm{Rest})$ is $O(\| K_{u}\|_{0}K^2)$ and the
inversion takes $O(K^3)$. Similarly, for updating the conditional
distribution of $\mathbf{v}_j$'s, the computational complexity is dominated
by $O(K^3)$. Recall for UPG the computational complexity for the E-step
for each user $u$ is $O(\| K_u\|_{0}^3 + LJ^2)$. While $\| K_u\|_{0}^3$
is generally smaller than $K^3$ in practical applications since a large
fraction of users interact with a small number of items, the additional
$O(LJ^2)$ term due to the conjugate gradient step adds considerable
complexity to UPG for large item sets. Hence, introducing sparsity
through Glasso that reduces $O(LJ^2)$ to almost linear in $J$ helps
with speeding up computation for the E-step in UPG. However,
introducing such sparsity comes at the cost of performing a Glasso in
the M-step. For small $J$ like in our PA application, UPG computations
are more scalable.

\section{Experiments and model comparions}
\label{secexp}
In this section we provide empirical analysis of our models with
comparisons to others.
We
report performance on two different data sets---(a) The benchmark
MovieLens 1M data set (described in Section \ref{secMlens}) from a
movie recommender system that has been
studied in the literature before, and (b) The Yahoo! PA data set described
earlier. For both data sets, we compare our \textit{UPG} models with several
existing methods.

%s6.1 ###
\subsection{Benchmark MovieLens 1M data} \label{secMlens}
We conducted experiments on\break a benchmark MovieLens 1M data set
(available at \href{http://www.grouplens.org}{www.grouplens.org}) that
consists of $1$M ratings
provided by 6,040 users on a set of 3,706 movies. The ratings
(response) $r_{uj}$ are on a 5-point
ordinal scale and the root mean squared error (RMSE) on out-of-sample
predictions has been used to evaluate different modeling methods on
this data before. Since reducing RMSE is the goal, statistical models
assume the response (ratings) to be Gaussian for this data. We sort
the training data by time stamp associated with each record and create a
$75\%\dvtx 25\%$ training-test split to evaluate performance. Note that in
this experiment we do not use any user or item covariates for any of
the models.
%User covariates include age, gender, zipcode (we used the first digit
%only) and occupation. The item covariates include movie genre.

%s6.1.1 ###
\subsubsection{Methods compared on MovieLens data}
We describe several collaborative filtering methods that are compared
to our approach. Some of these methods provide simple baselines, others
are state-of-the-art methods used in recommender problems.

\textit{Constant}---We assume $r_{uj} \sim N(\mu,\sigma^2)$ and predict
every rating to be a~constant $\mu$ estimated as the global mean of
training data.

\textit{Item--item similarity: IIS}---This is a classic model used in
recommender problems. In fact,
according to published sources [\citet
{linden03}; \citet{sarwar2001}], this
could be one of the key technologies behind Amazon's recommendation
engine. For the movie recommender problem, the rating of user $u$ on
item $j$ is predicted as
%
%e18 ###
\begin{equation}
\label{itemitem}
\bar{r}_{j} + \frac{\sum_{k \neq j}w_{jk}(r_{uk} - \bar
{r}_{k})}{\sum_{k \neq j}w_{jk}},
\end{equation}
where $w_{jk}$ measures similarity between items $j$ and $k$, and $\bar
{r}_k$ denotes the average rating on item $k$. For movie ratings data,
measuring $w_{jk}$ through
Pearson's correlation is popular [\citet{breese1998empirical}].

\textit{Most popular: MP}---We %augment LR to%
include both user and item main effects into the model. The
main-effects are treated as
random-effects and shrinkage is done using a normal prior. In other
words, we assume the following model for ratings $r_{uj}$:
\[
r_{uj} |(\alpha_u,\beta_j,\mu,\sigma^2) \sim N(\mu+\alpha_u+\beta
_j,\sigma^2),
\]
where $\alpha_u$ and $\beta_j$ are user and item random effects,
respectively, and $\mu$ is the global intercept.
The priors are given by
\[
\alpha_i \stackrel{\mathrm{i.i.d.}}{\sim} N(0,\sigma_{\alpha}^2);
\beta_j \stackrel{\mathrm{i.i.d.}}{\sim} N(0,\sigma_{\beta}^2).
\]

\textit{Bilinear random effects: BIRE}---We augment \textit{MP} to include
a multiplicative random effects term. In other words,
\begin{eqnarray*}
r_{uj} |(\alpha_u,\beta_j,\mu,\mathbf{q}_u,\mathbf{v}_j,\sigma^2) &\sim&
N(\mu+\alpha_u+\beta_j+\mathbf{q}_{u}^{\prime}\mathbf{v}_{j},\sigma^2),
\\
\alpha_u &\stackrel{\mathrm{i.i.d.}}{\sim}&
N(0,\sigma_{\alpha}^2); \qquad  \beta_j \stackrel{\mathrm{i.i.d.}}{\sim
} N(0,\sigma_{\beta}^2);
\\
\mathbf{q}_{u}
&\stackrel{\mathrm{i.i.d.}}{\sim}& \operatorname{MVN}(\mathbf{0},I);
\qquad
\mathbf
{v}_j \stackrel{\mathrm{i.i.d.}}{\sim} \operatorname{MVN}(\mathbf{0},aI).
\end{eqnarray*}
The inner product
of the $K$ dimensional user ($\mathbf{q}_u$'s) and item ($\mathbf
{v}_j$'s) random
effects, respectively, captures residual interaction. The
variance components for both \textit{MP} and \textit{BIRE} are
estimated by
fitting the
model through an EM algorithm. For \textit{BIRE}, we use an MCEM
algorithm [\citet
{rlfm}; \citet{salakhutdinov2008probabilistic}]. This model has
been extensively studied in the
literature and has been shown to provide state-of-the-art performance
compared to several other methods [\citet{koren2009matrix}].

%
%t1 ###
\begin{table}
\tabcolsep=0pt
\caption{Test-set RMSE on MovieLens Data. The number of factors for
\textit{BIRE} and the sparsity parameter $\rho$ for \textit{UPG}
reported obtained the best performance}
\label{tblrmse-movie}
\begin{tabular*}{\textwidth}{@{\extracolsep{\fill}}lcccccc@{}}
\hline
\textbf{Method} & \textbf{Constant} & \textbf{MP} & \textbf{IIS} & \textbf{BIRE} & \textbf{UPG} & \textbf{UPG-online} \\
&&&&\textbf{(15 factors)}&\textbf{($\bolds{\rho=0.002}$)}&\textbf{($\bolds{\rho=0.002}$)} \\
\hline
RMSE & 1.119 & 0.9643 & 0.9584 & 0.9435 & 0.9426 & 0.8733 \\ \hline
\end{tabular*}
\end{table}

%f7 ###
\begin{figure}[b]

\includegraphics{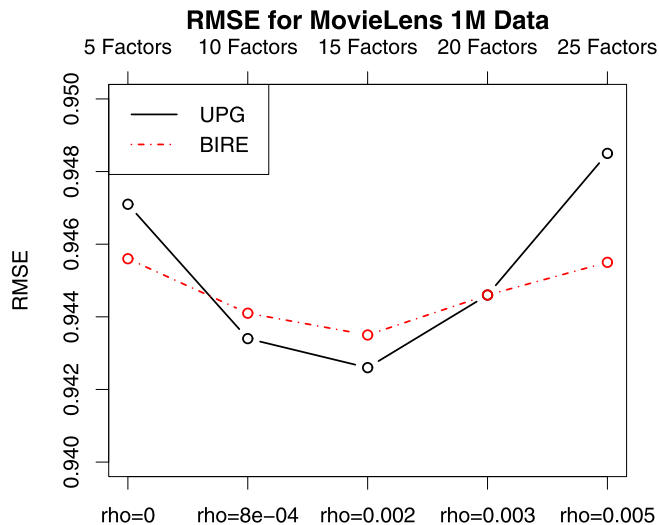}

\caption{For MovieLens 1M data, the RMSE performance of UPG model with
$rho$ equal to 0, 8\textup{e}--04, 0.002, 0.003, 0.005 compared to the BIRE model
with 5, 10, 15, 20 and 25 factors.}
\label{figUPGbire}
\end{figure}

\textit{UPG and UPG-online}---We fit the Gaussian version of our
\textit
{UPG} and \textit{UPG}-online models
for comparison to the methods described above. Naturally, for this
case, the PQL approximation is redundant.
\subsubsection{Discussion of results}
In Table~\ref{tblrmse-movie} we report the RMSE of various methods
on the $25\%$ held out test set. The \textit{Constant} model has poor
performance, and adding main effects to obtain \textit{MP} significantly
improves performance. The \textit{IIS} model is better than \textit{MP}
but significantly worse than \textit{BIRE}. Our \textit{UPG} model is
almost equivalent to \textit{BIRE} in performance. For both \textit{BIRE}
and \textit{UPG}, the hyper-parameters (number of factors for \textit
{BIRE}, sparsity parameter $\rho$ for \textit{UPG}) have an impact on
the RMSE performance. Figure~\ref{figUPGbire} shows the predictive
accuracy for different hyper-parameter values. In practice, these
parameters are estimated by cross-validation on a tuning set. Note
that for \textit{UPG} when $\rho=0$, the RMSE is worse than the
performance with $\rho=0.002$ (corresponds to 1.8\% nonzero diagonal
entries for the UPG model); this shows adding Glasso for precision
matrix regularization is important for large-item-set problems (we
have approximately $3.7K$ items in this case). To compare \textit{BIRE}
and \textit{UPG}, we analyzed the residuals from both these models on
the test data. Interestingly, the scatter plot revealed the residuals
to be strikingly similar with a Pearson's correlation of $0.97.$ For
this data, the more generic assumption for the precision matrix in
\textit{UPG} yields results that are similar to a low-rank
approximation provided by \textit{BIRE}.

Following a suggestion by a referee, we also compared our method with Fast
Maximum Margin Matrix Factorization (\textit{FMMMF}) [\citet
{rennie}] that predates
the \textit{BIRE} model. This approach also predicts ratings through the
multiplicative random
effects model as in \textit{BIRE} but replaces the Gaussian assumption
(squared error loss)
on the movie ratings by a~hinge loss function that incorporates the
ordinal nature of the
ratings. Due to nondifferentiability of the hinge
loss, it is approximated with a smooth hinge loss with Gaussian priors
on the user and item
factors as in \textit{BIRE}. The optimization is performed through
conjugate-gradient instead of the
MCEM algorithm as in \textit{BIRE}. As suggested in that work, we compare
performance in terms
of mean absolute error (MAE).\footnote{Performance of \textit{FMMMF} was
much worse than \textit{BIRE}
and \textit{UPG} in terms of RMSE, hence not reported.} The MAE of
\textit{FMMF} obtained after optimizing
all parameters through Matlab code available publicly from the authors
was   0.7997, compared to
  0.7077 achieved by \textit{UPG} ($\rho=0.002$).

The \textit{UPG}-online model leads to a significant improvement in
accuracy since
there are a large number of new users in the test set. For these users,
online updates to the posterior based on their prior ratings on items
help in obtaining better
posterior estimates of $\bolds{\phi}_u$'s and lead to more accurate
predictions of ratings. We computed the difference between mean
absolute residuals obtained from \textit{UPG}-online
versus \textit{UPG} normalized by the sample variance as done in the
standard two-sample $t$-test. The test statistic values
for users with zero observation and at least one observation in the
training set were 42.03 and 11.04, respectively.
Both groups had large sample sizes and the $p$-values from the $t$-test are
close to zero; this is suggestive of larger improvements through
\textit{UPG}-online for users with
no observations in the training set.

The practical significance of RMSE improvements on the
actual movie recommender system is hard to gauge. Large scale
recommender systems deployed in practice are complex and predictive
models are only one aspect (albeit an important one) that determine
quality. But to provide some idea, the RMSE differences of
top-50 entries in the recently concluded Netflix competition were
within $1\%$ of the winning entry.

%[Figure \ref{figUPGbire} about here]

For the best $\rho$ value which is 0.002, the \textit{UPG} model
gives a
precision matrix with around 1.8\% nonzero off-diagonal entries. The
sparsity of the precision matrix not only improves the RMSE
performance but also provides interpretable results in terms of item--item
conditional similarities. We analyze the estimated partial correlations
from the
\textit{UPG} model. For each pair of items $i$ and $j$, we consider the
partial correlation $\rho_{ij}$ [\citet{kendall1961advanced}] between
the random effects $\phi_{ui}$ and $\phi_{uj}$ defined as
%
%
%e19 ###
\begin{equation}
\rho_{ij}=\frac{-\Omega_{ij}}{\sqrt{\Omega_{ii}\Omega_{jj}}}.
\end{equation}
Intuitively, user preferences on two items $i$ and $j$ are associated
if $|\rho_{ij}|$ is large. If $\rho_{ij}=0$, then user random effects
for items $i$ and $j$ are conditionally independent. For MovieLens 1M
data, the top-10 movie pairs with the highest absolute values of
partial correlations are shown in Table
\ref{tablemoviepartialcorrelation}. Note that all pairs are sequels
and have positive partial correlation values. Also, if we look for the
highly related movies to a specific movie in the precision matrix,
for example, Toy Story (1995), we obtain movies such as Toy Story~2 (1999),
Mulan (1998), A Bug's Life (1998) and The Lion King (1994), etc.,
which are all cartoons.

%
%t2 ###
\begin{table}
\tabcolsep=0pt
\tablewidth=300pt
\caption{Pairs of movies with top 10 absolute values of partial
correlations in the precision matrix from \textit{UPG} $\rho=0.002$}
\label{tablemoviepartialcorrelation}
\begin{tabular*}{300pt}{@{\extracolsep{\fill}}lc@{}}
\hline
\textbf{The pair of movies} & \textbf{Partial correlation}\\
\hline
The Godfather (1972) &\\
The Godfather: Part II (1974) & 0.622\\
[3pt]
Grumpy Old Men (1993) &\\
Grumpier Old Men (1995) & 0.474\\
[3pt]
Patriot Games (1992) & \\
Clear and Present Danger (1994)& 0.448\\
[3pt]
The Wrong Trousers (1993) & \\
A Close Shave (1995) & 0.443\\
[3pt]
Toy Story (1995) &\\
Toy Story 2 (1999) & 0.428\\
[3pt]
Austin Powers: International Man of Mystery (1997) &\\
Austin Powers: The Spy Who Shagged Me (1999) & 0.422\\
[3pt]
Star Wars: Episode IV---A New Hope (1977) &\\
Star Wars: Episode V---The Empire Strikes Back (1980) & 0.417\\
[3pt]
Young Guns (1988) & \\
Young Guns II (1990) & 0.395\\
[3pt]
A Hard Day's Night (1964) & \\
Help! (1965) & 0.378\\
[3pt]
Lethal Weapon (1987) & \\
Lethal Weapon 2 (1989) & 0.364\\
\hline
\end{tabular*}
\end{table}

%to obtain better estimate of $w_{j,k}$ by assuming
%a decomposition $w_{j,k} = \mathbf{u}_{j}'\mathbf{u}_{k}$ and
%estimating the
%factors $\mathbf{u}$'s. We will not compare against this method since
%we
%compare against another method
%called matrix factorization that have been shown to be superior to all
%other existing methods in collaborative filtering. }

%s6.2 ###
\subsection{The Yahoo! PA data}
%s6.2.1 ###
\subsubsection{Methods compared}
We provide a detailed analysis of PA data with comparison to several
existing methods in the recommender literature. The methods compared
would differ slightly from those used for MovieLens 1M data due
to the binary nature of the response.
Since maximizing total clicks in a given time period is our goal, we
consider a metric that provides an
unbiased estimate of total clicks obtained for a set of visits.
To ensure unbiasedness, we compute this metric on a small fraction of
data that is obtained by randomly selecting visits on Yahoo! front page
and serving items at random for each of them. Obtaining such randomized
data with no
serving bias is unique and not typically available. We shall prove
how the metric computed on this data provides an unbiased estimate of
lift in click-rates. We begin by describing the comparative methods for
this data.

\textit{Per-item regression: IReg}---This is our per item logistic
regression model as described in Section~\ref{secmodel}.
User affinity to items is measured only through user covariates, we do
not consider a per-user model.

\textit{Item--Item Similarity: IIS}---The similarity measure $w_{jk}$ in
equation (\ref{itemitem}) is given by the Jaccard coefficient
[\citet
{jaccard1901etude}] which is
the fraction of users who click on both items $j$ and $k$ out of all
users who click on either items $j$ or $k$. The rating $r_{uj}$ in this
case is the click-rate.

\textit{Probabilistic Latent Semantic Indexing: PLSI}---\textit{PLSI}
was developed by Hof\-mann (\citeyear{hofmann1999probabilistic}) to model the joint
distribution of
user and item responses in collaborative filtering applications. It was
recently used for a news recommendation application
by Google [\citet{das2007google}]. The main idea is to use a mixture
model where, conditional
on the mixture memberships, user and items are independent. More
formally,
\[
p(j|u) = \sum_{l=1}^{K}p(j|l)p(l|u),
\]
where $l$ denotes
the latent membership variable and symbol $p$ denotes appropriate
distributions. Model fitting is conducted through an EM algorithm.

\textit{BIRE, UPG, UPG-online}---Other than the methods
described above, we also compare \textit{BIRE} with \textit{UPG} and
\textit{UPG}-online.

%

% { Metrics to evaluate performance} --- We begin by
%s6.2.2 ###
\subsubsection{Metrics to evaluate performance}
We begin by defining an estimator that provides an unbiased estimate of expected
total clicks in $T$ visits [see \citet{scavenging}]. To ensure
unbiasedness,
we collect data through a completely
randomized recommendation scheme. For a~small fraction of randomly
selected visits, we display two randomly selected items in the
recommended slots in region 2. Due to the randomization, the set of
visits that obtained a click is a random subsample. Confining
ourselves only to the clicked subsample on position 1, we measure the
number of times the model would have selected the clicked item. In
other words, we measure the concordance between the top-ranked items
selected by our statistical method and the ones that got clicked. More
specifically, we measure the performance of a model $M$ through the
measure defined as
%
%
%e20 ###
\begin{equation}
\label{eqnF1hit}
S(M) = J\sum_{\mathrm{visits\  with\ click}}1(\mathrm{item\ clicked} =
\mathrm{item\ selected\ by\ M}).
\end{equation}
We show $S(M)$ is an unbiased estimator of total clicks obtained by
serving items on position 1 through model $M.$
Let $c_{t,M(u_t)}$ denote the binary click random variable when item
$j_t=M(u_t)$ is served on visit $t$ to user $u_t$ by model
$M$($t=1,\ldots,T$).
Then, total expected clicks $V(M) = E(\sum_{t=1}^{T}c_{t,M(u_t)})=\sum
_{t=1}^{T}\sum_{j=1}^{J}E(c_{t,j}1(M(u_t)=j))=T\sum
_{j=1}^{J}E(c_{j}1(M(u)=j)),$ assuming
$(u_t,\break c_{t,1},\ldots,c_{t,J})$ are i.i.d. from some distribution. Now,
note that $S(M) =\break \sum_{t=1}^{T}c_{t,j_t}\times 1(M(u_t)=j_t)$, where $j_t$
is the item
selected by the randomized serving scheme for visit $t$:
%
%
%e22 ###
%e21 ###
\begin{eqnarray}
E(S(M)) &=& J\sum_{j=1}^{J}E\biggl(\sum_{t\dvtx j_t=j}c_{t,j}1\bigl(M(u_t)=j\bigr)\biggr)
\nonumber\\
&=& J\frac{T}{J}\sum_{j=1}^{J}E\bigl(c_{j}1\bigl(M(u)=j\bigr)\bigr)\\
&=& V(M).\nonumber
\end{eqnarray}
Note that the second inequality follows because $c_{t,j}$ is
independent of $u_t$ since we are evaluating on randomized data, and
since $u_t$'s are
random samples, $c_{t,j}1(M(u_t)=j)$ has the same distribution. Also,
$|t\dvtx j_t=j| = T/J$ under a~randomized serving scheme.
To compare the click-lift of model $M_1$ relative to $M_2$, we use
$100(\frac{S(M_1)}{S(M_2)}-1)$.
We report click-lift of various models under consideration relative to
the random serving scheme. Proof of unbiasedness under more
general scenarios is provided in \citet{scavenging}; we included
it for
the randomized data here for easy reference.\looseness=1

%s6.2.3 ###
\subsubsection{Discussion of results on PA data}
\begin{figure}

\includegraphics{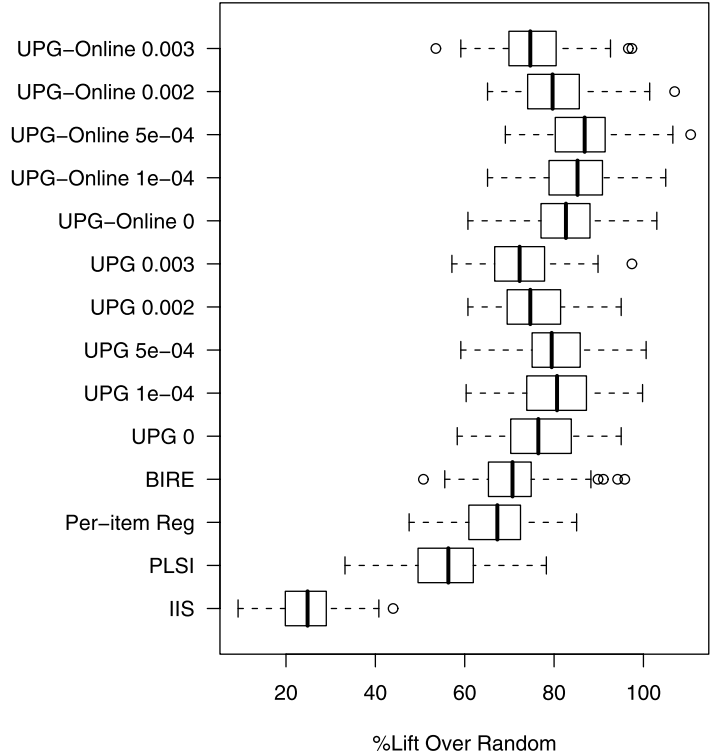}

\caption{For the PA data, the click-lift measure over random model for
the following models: Item--item similarity model (IIS), PLSI,
Per-application logistic regression model (IReg), BIRE with 15 factors,
UPG with $\rho=0$, UPG with $\rho=1\mathrm{e}$--$04$, UPG with $\rho=5\mathrm{e}$--$04$, UPG
with $\rho=0.002$, UPG with $\rho=0.003$, UPG-online with $\rho=0$,
UPG-online with $\rho=1\mathrm{e}$--$04$, UPG-online with $\rho=5\mathrm{e}$--$04$, UPG-online
with $\rho=0.002$, and UPG-online with $\rho=0.003$.}
\label{figf1-hit}
\end{figure}

We discuss results obtained for the click-lift measure over the random
model as shown in
Figure~\ref{figf1-hit}. The boxplot for each model represents the
performance of 20
bootstraps on the test data. We note that all models
achieve a significant click lift compared to the random model. The models
\textit{IIS} and \textit{PLSI} do not incorporate user covariates and
are worse than all others.
As expected, \textit{IReg} is better than \textit{IIS} and \textit{PLSI}
but \textit{BIRE} (15 factors) does better. However, for this
data set \textit{UPG} is significantly better than \textit{BIRE}.
Furthermore, the best \textit{UPG} model is based
on a nonzero Lasso penalty. As expected, \textit{UPG}-online provides
better results than \textit{UPG} offline.

%f8 ###

Collaborative filtering approaches that do not
incorporate covariates (\textit{IIS} and \textit{PLSI} in this case)
perform poorly. The per-item regression \textit{IReg} and state-of-the-art
\textit{BIRE} perform much better, but our new model that combines
covariates with item--item similarity in a model based way is
significantly better and achieves almost an $80\%$ improvement in
click-lift over random.

%s6.2.4 ###
\subsubsection{Interpretability of UPG for PA}
As in MovieLens, partial correlations for the \textit{UPG} model are
interpretable.
We report item pairs with top 10 absolute partial correlations in Table
\ref{tablepartialcorrelation}. Interestingly, all of
the partial correlations shown in the table are positive. The top two
pairs are \{``Fantasy
Sports,''\footnote{\url{http://sports.yahoo.com/fantasy}.}
``Fantasy
MLB''\footnote{\url{http://baseball.fantasysports.yahoo.com/}.}\},
\{``Fantasy Sports,'' ``Fantasy
Football''\footnote{\url
{http://football.fantasysports.yahoo.com/}.}\}. Note
that ``Fantasy MLB'' and ``Fantasy Football'' are two different kinds
of fantasy
games in ``Fantasy Sports.'' The 3rd ranked pair \{``AOL
Mail,''\footnote{\url{http://webmail.aol.com}.} ``Gmail''\} are very
well-known email services provided by AOL and Google, respectively. The
4th ranked pair \{``PEOPLE.com,'' ``EW.com Featured''\footnote{Top
stories in
\url{http://www.ew.com/ew}.}\} are both news site on celebrities
and entertainment.

%
%t3 ###
\begin{table}
\tabcolsep=0pt
\tablewidth=240pt
\caption{Pairs of items with top 10 absolute values of partial
correlations in the dense precision matrix from \textit{UPG} ($\rho=0$)}
\label{tablepartialcorrelation}
\begin{tabular*}{240pt}{@{\extracolsep{\fill}}lcc@{}}
\hline
\textbf{Item 1} & \textbf{Item 2} & \textbf{Partial correlation}\\
\hline
Fantasy Sports & Fantasy MLB & 0.556\\
Fantasy Sports & Fantasy Football & 0.434\\
AOL Mail & Gmail & 0.367\\
PEOPLE.com & EW.com Featured & 0.265\\
Shopping & Personals & 0.237\\
PEOPLE.com & PopSugar & 0.224\\
Travel & Shopping & 0.222\\
News & Shopping & 0.208\\
EW.com Featured & PopSugar & 0.182\\
News & Personals & 0.181\\
\hline
\end{tabular*}
\vspace*{3pt}
\end{table}

%s6.3 ###
\subsection{Timing comparison between UPG and BIRE}\label{sectimes}
For MovieLens 1M data and Yahoo! PA data, we compare the time required
for training the \textit{UPG} models and \textit{BIRE}.
Section~\ref{secupgcomp} describes the computational complexity of the
\textit{UPG}.

The timing comparison between \textit{BIRE}
and \textit{UPG} with different sparsity parameter $\rho$ is shown in
Table \ref{tabletimingmovielens}. The M-step for \textit{UPG}
includes only the timing for Glasso, the computation of the sample covariance
being included in the E-step.
Based on the table, we see that
\textit{UPG} takes more time than \textit{BIRE}. Following our
discussions in
Section \ref{seccompcomplexbire}, this is
because of the (major) time spent in updating
$\sum_u {\bolds\Sigma_u}$ and $\mu_u$, which is large due to the size
of $J$.
The timings of both E-steps and M-steps decreased with increasing
sparsity level, as expected from our prior discussions.
The E-step and M-step timings are not directly comparable, since the
E-step uses multiple threading, where the M-step does not.

For relatively low dimensional problems (small $J$) but a large number
of users, \textit{UPG}
could be much \textit{faster} than \textit{BIRE}, which is shown in the case
of the Yahoo! PA data
set. For 140K users and 51 items, \textit{UPG} takes only 7 seconds per
iteration using 7 threads, while \textit{BIRE} with 15 factors and 100 samples
per E-step takes 3378.1 seconds per iteration.
With larger $\rho$ and increased sparsity, the timings do improve
slightly, but the improvement is not as
prominent as in the MovieLens example (with larger number of items).

Therefore, for many real-life problems such
as Yahoo! PA, the \textit{UPG} models can have significant edge over
\textit{BIRE} both
in terms of accuracy and computation time.

%
%t4 ###
\begin{table}
\tabcolsep=0pt
\caption{For MovieLens 1M data, the timing comparison (s) between
\textit{BIRE} and \textit{UPG} (7 threads) with different sparsity parameter
($\rho$) values. BIRE uses 15 factors and 100 MCMC samples per E-step,
single thread. Note that ``0'' means time is negligible because both
BIRE and UPG ($\rho=0$)~do~not~use Glasso in the M-step}
\label{tabletimingmovielens}
\begin{tabular*}{\textwidth}{@{\extracolsep{\fill}}lcccccc@{}}
\hline
& \textbf{BIRE} & \textbf{UPG} & \textbf{UPG} & \textbf{UPG} & \textbf{UPG} & \textbf{UPG}\\
& \textbf{15 factors} & $\bolds{\rho=0}$ & $\bolds{\rho=8\mathrm{e}$--$04}$ & $\bolds{\rho=0.002}$ & $\bolds{\rho
=0.003}$ & $\bolds{\rho=0.005}$\\
\hline
E-Step & 208.9 (s) & 6555.0 (s) & 5254.7 (s) & 4049.7 (s) & 3466.9 (s)
& 2857.0 (s) \\
M-Step & 0 & 0 & 3502.3 (s) & 2405.5 (s) & 1964.3 (s) & 1622.9 (s)\\
\% Nonzero  off-& NA & 100\% & 7.3\% & 1.8\% & 0.9\% &
0.3\% \\
 \quad diagonals of $\bolds{\Omega}$\\
\hline
\end{tabular*}
\end{table}

%s7 ###
\section{Discussion}
\label{secdisc}
Although not widely studied in the statistics literature, the problem
of algorithmically recommending items to users has received a~lot of
attention in computer science and machine learning in the
last decade. Large scale recommendation systems are mostly operated by big
organizations like Amazon, Netflix, Yahoo! and Google; sharing data
for academic research is difficult due to privacy concerns. A
significant breakthrough was achieved when Netflix
decided to run a competition and released a large amount of movie
ratings data to the public in October, 2006.
Since then, several methods have been published in the academic
literature. Of these, the \textit{BIRE} model
described earlier has emerged to be the best, other classical methods
like item--item similarity
are not as accurate. Methods based on classical and successful data
mining techniques like restricted Boltzmann machines (\textit{RBM})
were also tried, but they were comparable and in some cases worse than
\textit{BIRE} [\citet{Salakhutdinov2007RBM}].
However, the errors in predictions made by \textit{BIRE} and \textit
{RBM} were found to be uncorrelated on Netflix, hence, an ensemble approach
that combined \textit{BIRE}, \textit{RBM} and several other methods
eventually won the competition.

In this paper we proposed a new hierarchical model called \textit{UPG}
that generalizes \textit{BIRE}
by moving away from modeling item--item similarity in terms of a low
rank matrix. Through extensive analysis (one
benchmark data and a new data set from Yahoo! front page) we show that
our approach is significantly better in one application (Yahoo! PA data)
and comparable to \textit{BIRE} on the benchmark data set. In fact,
comparing residuals obtained from \textit{UPG} and \textit{BIRE} on
the MovieLens data, we found them
to be highly correlated (Pearson's correaltion coefficient was about
$ 0.97$!). Thus, even in applications where a low-rank approximation
does provide
a good model, the performance of our \textit{UPG} model is comparable.
In applications like PA where a low-rank approximation is not suitable,
the flexibility
of \textit{UPG} leads to better accuracy. Other than accuracy,
\textit{UPG} also provides interpretable results in terms of item--item
similarity. In many practical applications, item--item
methods are still used since they provide interpretable results. Our
\textit{UPG} model has both features---accuracy and interpretability.
We believe this makes it a promising method that could potentially lead
to interesting future research in this area.

The objective of the PA recommender problem described in this paper is to
maximize the total clicks on the module in some long time horizon. This
is a bandit
problem since there is positive utility associated with taking risk
and showing items with low mean but high
variance [\citet{bookberry85}]. There exists a rich literature on
bandits that is related to this
problem [\citet{mljauer02}; \citet{lai85};
\citet{assarkar91}; \citet
{tacwang05}; \citet{asyang02}].
However, bandit
algorithms without dimension reduction
may converge slowly in high-dimensional problems. Our \textit{UPG}
models provide a possible way to achieve such dimension reduction by exploiting
correlations among response for different items; this reduces the
amount of exploration required
to perform personalization at the user level. Since we work with a
well defined statistical model, it is possible to obtain both mean and
uncertainty estimates. These can be coupled with bandit schemes like
UCB [\citet{mljauer02}]
to obtain faster convergence to the best item per user. We can also use
the mean
estimates alone with bandit schemes that do not require explicit
variance estimates
from statistical models [\citet{focsauer95}]. In our PA
application, we
found simple $\epsilon$-greedy
to work well since the number of items in the pool was small and sample
size available on Yahoo! front page is large.
In other scenarios where sample size available per item is small
due to large item pool and/or item churn, other bandit schemes
like UCB may perform significantly better and can be easily coupled
with the output of our \textit{UPG} models. Constructing a bandit
scheme that is optimal for our \textit{UPG} model
is more involved and is an example of correlated bandits. Some works on
correlated bandits exist in the
literature, but they do not directly apply to our \textit{UPG} model
[\citet{Kleinberg2008}; \citet
{icmlsandeep07}; \citet{srinivascorr}].
We leave the investigation of optimal bandit schemes for \textit{UPG}
as future work. Similar comments apply to \textit{BIRE} for which
constructing optimal bandit schemes has not been investigated and is
still an open problem.

Several other open issues remain to be addressed carefully. Although we
were able to scale our
method to approximately $4$K items, scaling to larger item sets
requires more research. As we pointed out in Sections~\ref{secupgcomp}
and~\ref{seccompcomplexbire}, the main computational
bottlenecks for \textit{UPG} are the $O(J^2)$ conjugate gradient
computations in the E-step and $O(J^3)$ computation for Glasso.
Sparsity helps significantly but more research is required. For scaling
up the E-step, more large scale distributed
computation can help (we only used $7$ threads in this paper). However,
the Glasso algorithm needs more work with large~$J$.
For instance, in our current implementation we optimize each
intermediate graphical-Lasso till moderate/high accuracy.
Since partial optimization of the convex objective in the M-step is
good enough in the early iterations of our EM procedure,
this strategy deserves further investigation.
We tried to modify the Glasso algorithm of \citet{friedman2008sparse}
along these lines,
but we observed that the positive-definiteness of the returned
precision matrix was not always preserved, thus our
algorithm [\citet{prox-glasso-2011}] for this purpose was more favorable.
Other than improving optimization, it is also worthwhile to scale the
computations by using model approximations. One
possibility is to cluster the item set into a
smaller number of categories and capture item--item dependencies through
category-category associations. Such a compression, though scalable, may
lead to a poor fit and hurt prediction accuracy. How to perform such
clustering to achieve a good
compromise between scalability and accuracy is an interesting
question. Also, in several web recommender problems, there is
significant item churn over
time. We are currently investigating methods that generalize our
methodology to update the similarity matrix
for such applications. Overall, we believe our \textit{UPG} model
opens up new research avenues
to study problems arising in the area of recommender problems.

\section*{Acknowledgments}
The authors would like to thank the anonymous reviewers and the Editor
for their helpful comments and suggestions
that
improved the presentation of the paper.
We would also like to thank Trevor Hastie from
Stanford University for helpful discussions.
Rahul Mazumder would like to thank the hospitality
of Yahoo! Research for the summer of 2010,
during which he got involved in the work.

% AOS,AOAS: If there are supplements please fill:
% \sname{Supplement A}
% \stitle{Title}
% \slink[url]{http://lib.stat.cmu.edu/aoas/???/???}
% \sdescription{Some text}

%suskaldyti doi

% imsref loaded by smiklovaite, 2011-05-13 15:22:45
% imsref loaded by smiklovaite, 2011-05-13 16:03:37
%

\printaddresses

\end{document}